\documentclass[
 preprint,
groupedaddress,
 amsmath,amssymb,
aps,
]{revtex4-1}
\usepackage{multirow}
\usepackage{hyperref}
\usepackage{color}
\usepackage{textcomp}
\usepackage{amsthm}
\usepackage{helvet}
\usepackage[T1]{fontenc}
\usepackage{times}
\usepackage{bm}
\usepackage{CJK}
\usepackage{eso-pic}
\usepackage{wasysym}
\usepackage{subfigure}
\usepackage{graphicx}
\usepackage{dcolumn}
\usepackage{bm}
\usepackage[colorinlistoftodos]{todonotes}

\begin{document}

\preprint{APS/123-QED}

\title{Quantum oscillations from networked topological interfaces in a Weyl semimetal}

\author{
I-Lin Liu,$^{1,2,3}$
Colin Heikes,$^1$
Taner Yildirim,$^1$ 
Chris Eckberg,$^2$
Tristin Metz,$^2$
Sheng Ran,$^{1,2,3}$
William Ratcliff II,$^1$
Johnpierre Paglione,$^{2}$ 
and Nicholas P. Butch$^{1,2}$}
 
\affiliation{\vspace{4pt}
$^1$NIST center for Neutron Research, NIST, Gaithersburg, Maryland 20899, USA.
\\$^2$Center for Nanophysics and Advanced Materials, Department of Physics,  University of Maryland, College Park, MD 20742 USA.
\\$^3$Department of  Materials Science and Engineering, University of Maryland, College Park, MD 20742 USA.}

\date{\today}





\begin{abstract}
Layered transition metal chalcogenides are promising hosts of electronic Weyl nodes and topological superconductivity\cite{Qian2014,Qi2016,Hsu2017}. MoTe$_2$ is a striking example that harbors both noncentrosymmetric T$_d$ and centrosymmetric T' phases, both of which have been identified as topologically nontrivial \cite{Soluyanov2015,Deng2016,Huang2016,Jiang2017,HOTIBernevig}. Applied pressure tunes the structural transition separating these phases to zero temperature\cite{Heikes2018}, stabilizing a mixed T$_d$-T' matrix that entails a unique network of interfaces between the two non-trivial topological phases. Here, we show that this critical pressure range is characterized by unique coherent quantum oscillations, indicating that the change in topology between two phases give rise to a new topological interface state. A rare combination of topologically nontrivial electronic structures and locked-in transformation barriers leads to this counterintuitive situation wherein quantum oscillations can be observed in a structurally inhomogeneous material. These results open the possibility of stabilizing multiple topological superconducting phases, which are important for solving the decoherence problem in quantum computers.

\begin{description}
\item[PACS numbers]
May be entered using the \verb+\pacs{#1}+ command.
\end{description}
\end{abstract}

\pacs{Valid PACS appear here}
\maketitle

Topologically protected electronic states at material interfaces are attractive because they cannot be destroyed by many types of perturbations\cite{Kitaev2003,Nayak2008}. Topological superconductivity is such a protected quantum state, stable to local noise and disorder, that is being considered as a platform for decoherence-free, universal quantum computation\cite{Kitaev2003,Nayak2008,Qizhang2011}.
An especially fruitful host of this exotic state is MoTe$_2$, which has had both its bulk orthorhombic T$_ d$ phase, and hole-doped monolayer specimens identified as possible topological superconductors\cite{Qian2014,Qi2016,Hsu2017}. In addition, a topological superconducting phase was recently discovered in sulfur-substituted samples, with novel S$_{+-}$-wave pairing\cite{LiYanan2018}. These unusual superconductors all emerge from topologically nontrivial normal states: the T$_d$ phase has been identified as a type II Weyl semimetal\cite{Soluyanov2015,Deng2016,Huang2016,Jiang2017}, whereas the monoclinic T' phase is predicted to be a higher-order topological material \cite{HOTIBernevig}. In this work, we demonstrate experimentally how pressure drives MoTe$_2$ into three different regimes having nontrivial electronic topology\cite{Murakawa2013}, all of which host superconductivity. These nontrivial states are particularly robust and survive under significant structural disorder.

\begin{figure}
\centering
\subfigure{\label{PDr}\includegraphics[scale=0.46, trim=10 100 40 20,clip]{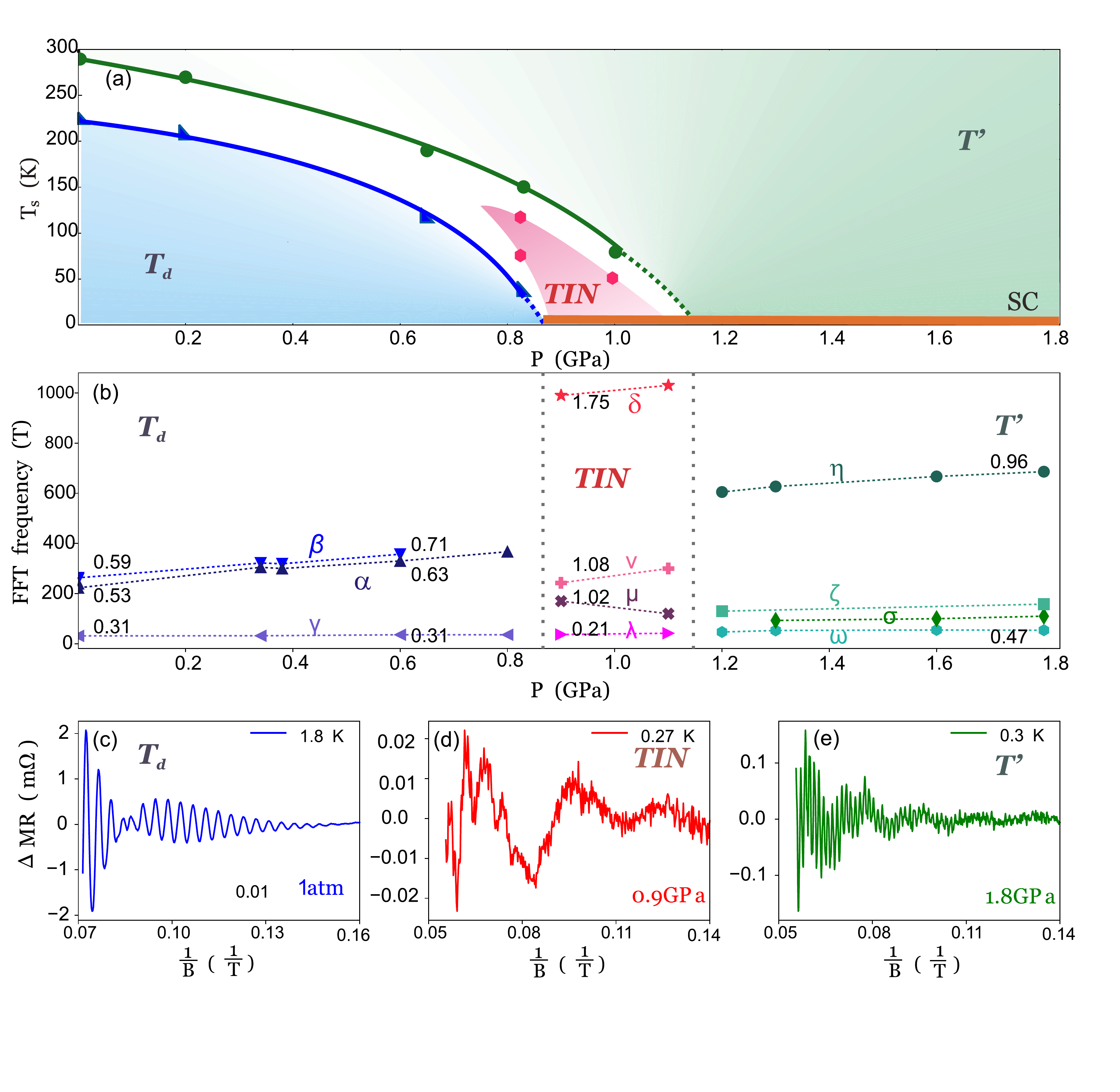}}
\subfigure{\label{pd-fft}}
\subfigure{\label{pd-1atm}}
\subfigure{\label{pd-1gpa}}
\subfigure{\label{pd-1p8}}
  \caption{(a) Pressure-temperature phase diagram of  MoTe$_2$. The superconducting temperature increases with pressure and turns into a full volumn superconducting transition and T' to T$_d$ Structural transition is totally suppressed above 0.8 GPa. The green diamonds and blue dots are the phase boundary of 100 $\% $ of T' and T$_d$ through neutron scattering. (b) Pressure dependence of FFT spectrum of oscillatory magnetoresistance. The numbers index the effective mass close to their markers under different pressure. The quantum oscillations with most pronounced pressure dependence, $\alpha$ and $\beta$ in the T$_d$ phase, and $\eta$ in the T' phase, correspond to  extremal orbits on the large electron pockets. In the TIN region, these disappear and are replaced by a completely new set of oscillations arising from topological interface states. The representative SdH oscillations of MoTe$_2$ recorded (c) at ambient pressure (Weyl structure in T$_d$), (d) 0.9 GPa (TIN), and (e) 1.8 GPa (higher-order topology in T'). Clear changes in the quantum oscillations reflect significant changes in the electronic structure.
  }
\end{figure}

The first-order structural transition separating the T' and T$_d$ phases in MoTe$_2$ has a distinct pressure dependence (Fig.~\ref{PDr}). At ambient pressure, the inversion-symmetric T' phase is stable at room temperature, only transforming into the noncentrosymmetric T$_d$ phase when cooled below roughly 250 K\cite{Qi2016,Heikes2018}. Neutron diffraction allows the determination of the relative volume fraction of these phases under different conditions\cite{Heikes2018}. As pressure increases, the transition temperature decreases. At pressures higher than 0.8 GPa, a new phenomenon emerges, where a roughly balanced mixture of the T' and T$_d$ phases stabilizes over an appreciable temperature range, and crucially, extends to the lowest measured temperatures. The existence of this unique frozen mixed-phase region is stabilized by the lack of sufficient entropy at these suppressed temperatures for atoms to move to their lowest-energy configuration, implying that there is a dominant extrinsic transformation energy barrier between two energetically nearly-degenerate structures \cite{Heikes2018}. 

\begin{figure}
\centering
\subfigure{\label{p1atm}\includegraphics[scale=0.52, trim= 190 390 210 420,clip]{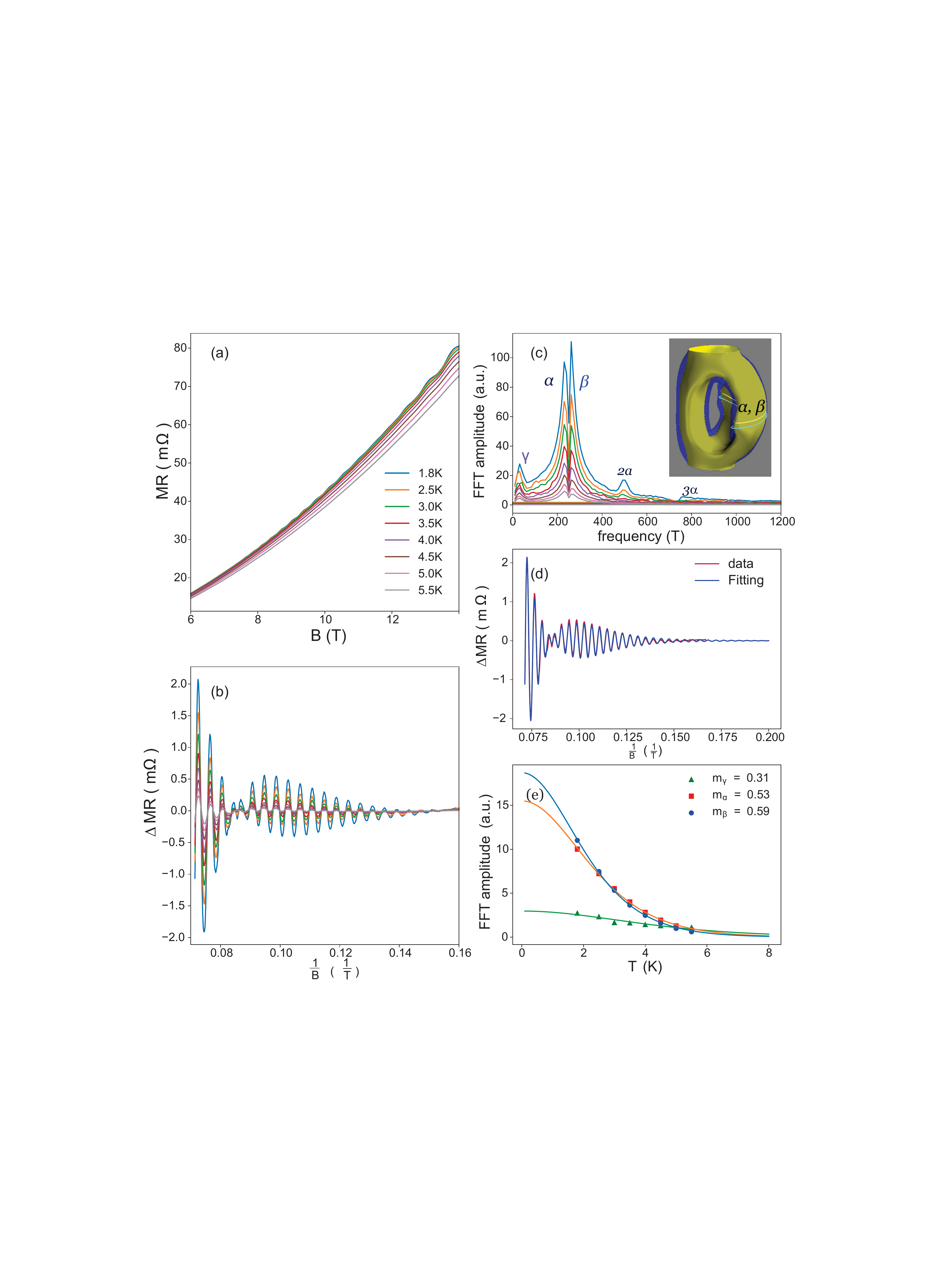}}
\subfigure{\label{p1atm-b}}
\subfigure{\label{p1atm-c}}
\subfigure{\label{p1atm-d}}
\subfigure{\label{p1atm-e}}
\subfigure{\label{p1atm-f}}
  \caption{(a) The longitudinal MR of the bulk T$_{d}$-MoTe$_{2}$ measured at ambient pressure with magnetic field parallel to c axis. (b) The corresponding SdH oscillations were observed by second-order polynomial background subtraction of normal magnetroresistance. (c) The FFT spectrum show three Fermi surfaces with oscillation frequencies at $F_{\gamma}$ = 32.5 T, $F_{\alpha}$ = 240.5 T and $F_{\beta}$ = 258 T. (d) Best fitting of SdH oscillation at 1.8 K and the related Berry's phases are nontrivial Berry phase, $\phi_{\gamma}$ = $\pi$, $\phi_{\alpha}$ = 0.88$\pi$, and $\phi_{\beta}$ = 0.88$\pi$. The 3D topological electron pockets of $\alpha$ and $\beta$ are supported by their 3D topological phase shift, $-\frac{1}{8}\pi$, nontrivival Berry's phases, and angle dependence SdH in supplementary info, 
and consistent with the observation of ARPES\cite{Deng2016}. (e) The effective mass are obtained by the temperature dependence of LK fitting.}
\end{figure}

The basic components underlying the Weyl semimetallic state of the T$_d$ phase are a large hole pocket centered on the Brillouin zone and two neighboring electron pockets along the $\Gamma - X$ direction\cite{Deng2016,Huang2016,Jiang2017,Crepaldi2017}. The hole pocket is observed in angle-resolved photoemission spectroscopy (ARPES)\cite{Deng2016,Huang2016,Jiang2017}, but is not apparent in SdH measurements\cite{Rhodes2017}. Prominent quantum oscillations observed in the T$_d$ phase arise from orbits associated with the electron pocket \cite{Rhodes2017}. Fig.~\ref{p1atm} and Fig.\ref{p1atm-b} show magnetoresistance and SdH oscillations at ambient pressure, in which these are clearly seen. As the Fast Fourier Transform (FFT) explicitly shows (Fig.~\ref{p1atm-c}), the beating seen in Fig.~\ref{p1atm-b} is due to two similar frequencies, $F_{\alpha}$ = 240.5 T and $F_{\beta}$ = 258 T, the result of symmetry-allowed spin-orbit splitting. First-principles calculations identify these frequencies with the larger extremal $k_z=0$ cross section of the electron pocket.

\begin{figure}{\label{FFT-Td}}
\centering
\subfigure{\includegraphics[scale=0.82, trim= 100 10 10 10,clip]{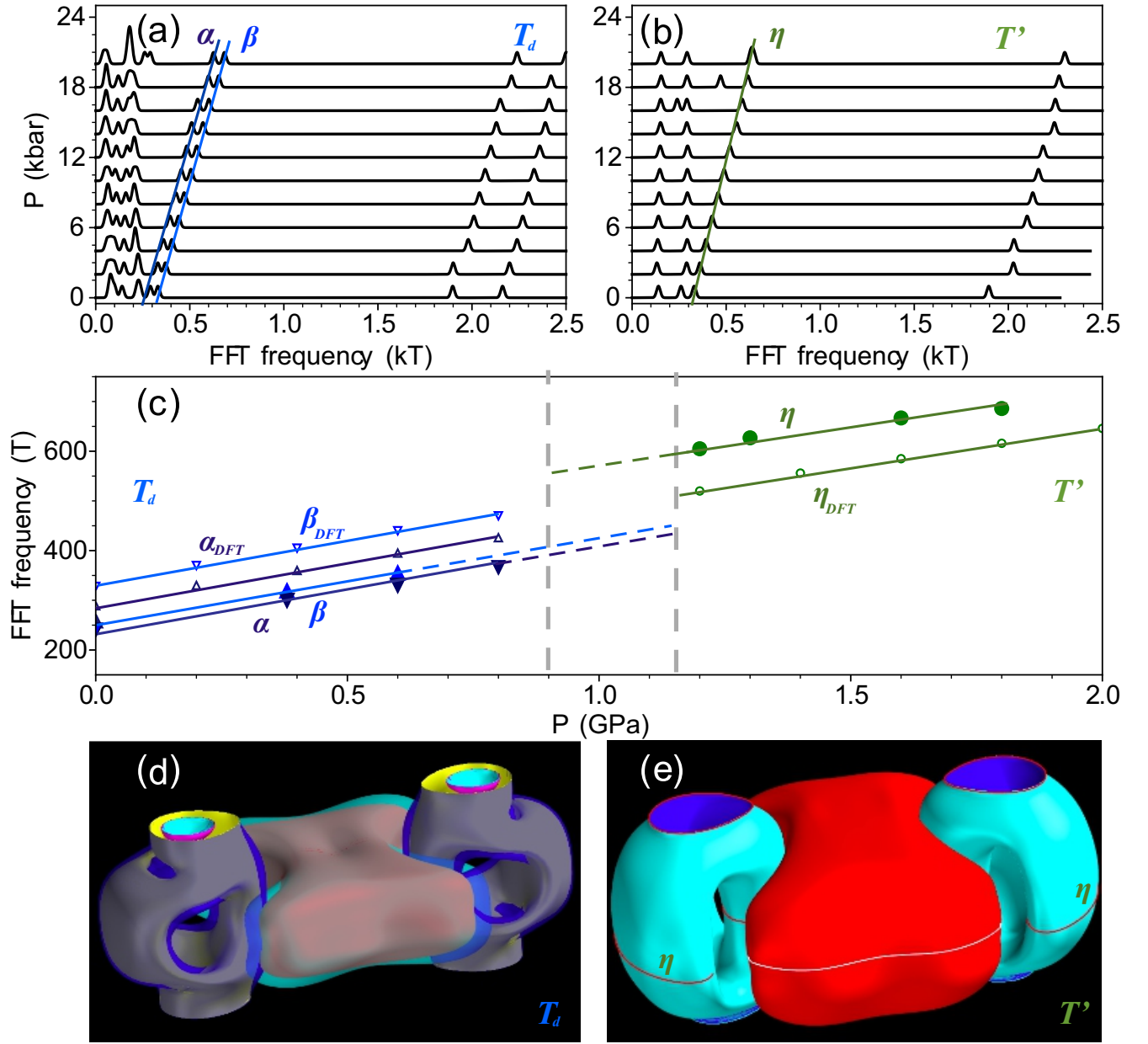}}{\label{FFT-T}}
\subfigure{\label{FFTcomparison}}
\subfigure{\label{Fermi-Td}}
\subfigure{\label{Fermi-T}}
    \caption{ (a,b) Calculated in-plane quantum oscillation frequencies for the Td and T' phases, showing that the SdH oscillations arising from the electron pockets, increase with increasing pressure in both phases. (c) A comparison between calculated and measured frequencies shows excellent quantitative agreement. (d, e) Calculated Fermi Surfaces of the Td and T' phases}
\end{figure}

Modeling of the SdH oscillations yields a remarkably good fit (Fig.~\ref{p1atm-d}) to the experimental SdH by Bumps (global fitting using Markov Chain Monte Carlo sampling\cite{bumps}), allowing the reliable determination of the Berry's phase for each frequency (Supplementary Table 1). Notably, all of the oscillations feature a nontrivial Berry's phase\cite{Murakawa2013}, consistent with a Weyl topology (see Supplementary). The effective band masses are light, and slightly less than previously reported \cite{Rhodes2017,Luo2016} as shown in Fig.~\ref{p1atm-f}. As a function of pressure, the electron pockets increase modestly in size due to lattice compression, but the nontrivial phase shift is maintained throughout the T$_d$ phase. This trend is consistent with first-principles calculations (Fig.~\ref{FFTcomparison}).


Throughout the T' phase, instead of two frequencies, a single frequency $F_\eta$ is observed, increasing from 600 T to 700 T over the measured range of pressure (Fig.~\ref{pd-fft}). The Fermi surface in the T' phase closely resembles that of the T$_d$ phase, although the symmetry of the T' phase removes the spin-orbit splitting of the bands contributing to the electron pocket (Fig.\ref{FFTcomparison}). Consistent with this, the pressure dependence of the electron pocket is similar in both T$_d$ and T' phases, and the calculated values of $F_\eta$ extrapolate from the $F_\alpha$, $F_\beta$ pair in the T$_d$ phase, quantitatively in good agreement with the measured frequencies. The T' phase has been predicted to harbor an unusual type of nontrivial topological state\cite{HOTIBernevig}. Although ARPES is ambiguous about whether the T' phase is topologically nontrivial \cite{Crepaldi2017}, our results strongly support a nontrivial topology. It is clear in our SdH oscillations that a nontrivial $\pi$ Berry's phase exists also in the T' phase (Supplementary Table 1). This has the exciting immediate implication that superconductivity in the T' phase may be inherently topologically nontrivial.

\begin{figure}{\label{compare}}
\centering
\subfigure{\label{pevop6}\includegraphics[scale=0.52, trim= 5 0 0 0,clip]{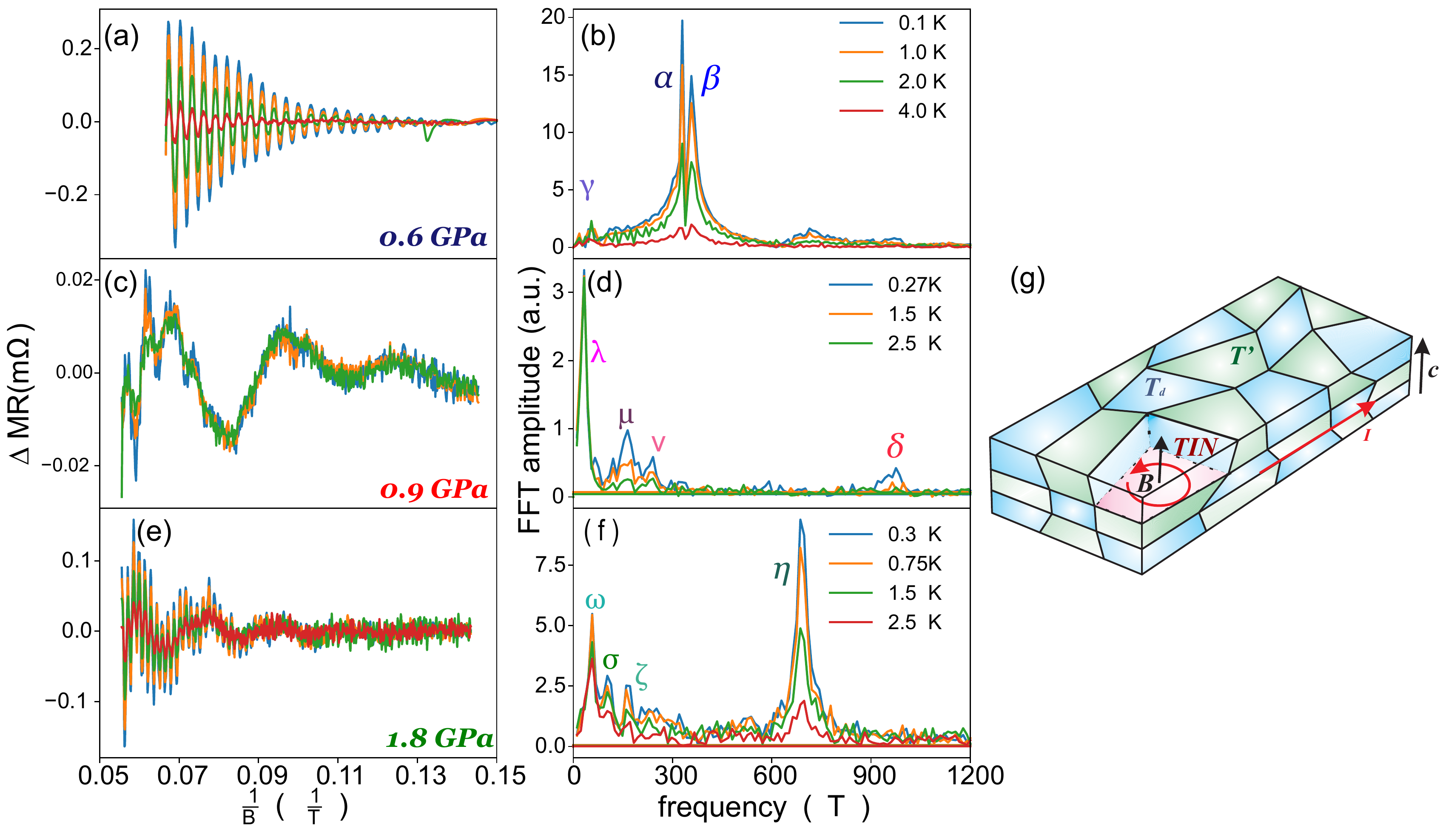}}
\subfigure{\label{fft-p6a}}
\subfigure{\label{sdh-p9}}
\subfigure{\label{fft-p9}}
\subfigure{\label{sdh-1p8}}
\subfigure{\label{fft-1p8}}
\subfigure{\label{model}}
  \caption{Temperature dependence of the SdH oscillations of MoTe$_2$ measured at (a) 0.6 GPa, (c) 0.9 GPa, and (e) 1.8 GPa represent for T$_{d}$-MoTe$_2$, TIN, and T'-MoTe$_{2}$. The amplitudes of SdH oscillations gradually decrease with increasing of pressure in T$_d$ and suddenly drop one order magnitude in surface state. In T', the amplitude increases up to one order magnitude. FFT spectrum at (c) T$_d$ and (i) T' shows that cross-section of their Fermi surfaces expand with increasing of pressure. The TIN (f) at 0.9 GPa suddenly prompt and discontinuity of the much weaker oscillation amplitudes, unique band structure with nontrivial Berry's phases and effective masses support the novel topological surface state different from bulk T$_{d}$ and T'. (g) Schematic of natural topological interface network (TIN). The quantum oscillation of bulk T$_d$ and T' were limited within the local grain boundaries and suppressed by the disorders. The networked interfaces between multilayer of two topological structures as Weyl semimetal and higher-order of topology can effectively screen the interaction between adjacent layers of different structures of transition metal dichalcogenide MoTe2, inducing the novel topologically protected edge states. The relative weak and coherent signals are robust under the small system perturbation and only come from the connected interfaces. This TIN heterostructure can parametrically increase the number of edge channels and is a promising approach to increasing the surface-to-volume ratio of mixed-phase topological materials. We only get SdH signal along ab plane since field is along c axis. The random grain boundary except ab plane can not get the coherent SdH oscillation signals.}
\end{figure}

The mixed region exists in a range of pressures and temperatures between the bulk T$_d$ and T' phases (Fig.~\ref{PDr}). It consists of an approximately balanced partial volume fraction of each. 
No other structural phases or ordered superstructures are apparent from neutron diffraction measurements. Naively, it might be expected that any measured SdH oscillations in the mixed region would consist of a superposition of T$_d$ and T' signals. However, we do not observe either. A natural explanation is that increased scattering typically weakens quantum oscillations. The mixed region is heterogeneous and sufficiently disordered that SdH oscillations from both the T$_d$ and T' phases are suppressed. In light of this, it is completely unexpected that a new set of unique SdH oscillations appears (Fig.~\ref{pevop6}). Furthermore, the oscillations in the mixed phase are reproducible after increasing and decreasing applied pressures through the critical range, confirming its intrinsic nature. Aside from the presence of new frequencies corresponding to new Fermi surfaces, a new band structure in the mixed region is inferred from a change in effective mass and much weaker oscillation amplitude relative to bulk T$_d$ and T', as shown in Fig.\ref{fft-p9}. In addition to the survival of these electronic states in the presence of disorder, the SdH oscillations from the mixed region feature nontrivial Berry's phases (Supplementary Table 1), features typical of a topologically protected state. This explains why no new structural phases characterize the mixed region - the topological states are surface states of the bulk phases, in this case, their interfaces. Thus, the mixed region serves as the foundation for a new type of electronic system in MoTe$_2$: a natural topological interface network (TIN),

Essential to the TIN state is the coexistence of two bulk phases with different topological invariants. The mixed region provides a natural framework for this coexistence. As the T$_d$ phase is a Weyl semimetal, its surface Fermi arcs have been much studied \cite{Deng2016} . However, the TIN differs potentially in one key aspect, namely, that unlike the vacuum, which is topologically trivial, the T' phase is topologically nontrivial. In order for edge states to exist at the interfaces, there must be a change in topology between T$_d$ and T' phase. No calculations exist to describe this interface, and in general, interfaces between two different topological bulk states have not received much theoretical attention. Our discovery suggests that this is a rich area for future exploration and exploitation.

In the TIN, due to the layered structure of both T$_d$ and T' phases, the $ab$ plane is preserved, and the largest grain boundaries fall along the $ab$ plane, which is the orientation probed by the SdH measurements ( Fig.\ref{model}). In MoTe$_2$, a naturally generated heterostructure provides an interesting demonstration of topological transport protection. The lateral dimensions of the interfaces are the same as those of the bulk grains, based upon which one naively expects similar damping of the SdH oscillations from the interfaces. Yet the clear SdH oscillations from the interfaces prove that the interfacial states have lower scattering than the bulk, and are a sign of their topologically nontrivial nature. In other words, the interfacial signal has been amplified by suppressing the bulk SdH oscillations through grain boundary scattering, and increasing the interface volume.

The pressure-tuned progression of different topological states raises the exciting possibility of studying several different types of topological superconductivity. Whereas superconductivity of the T$_d$ Weyl semimetal has received most attention, it is filamentary, and it has only recently been realized that bulk superconductivity exists in the T' phase. The topological classification of this superconducting state has not yet been explored. Additionally, the exact nature of the interfacial electronic states is a rich new direction of study, including possible new classes of superconductor. As we have demonstrated that the TIN can be readily stabilized, we can look forward to making use of these states in future topological quantum computation schemes and other applications.


\section{Methods}

{\bf First-principles Calculations. } The total energy, structure optimization under pressure, and band structure and Fermi surface calculations were performed by Quantum Espresso\cite{qe}, which is based on density-functional theory (DFT), using a plane wave basis set and fully relativistic all-electron projected augmented wave (PAW) potentials\cite{paw1,paw2}. The $4s, 4p, 4d$, and $5s$ electrons of Mo and the $4d, 5s$, and $5p$ electrons of Te were treated as valence. We used 0.02 Ry Methfessel-Paxton smearing with wavefunction and charge density cut-off energies of 100 Ry and 800 Ry, respectively. The exchange-correlation interactions were described by the generalized gradient approximation (GGA) with the Perdew-Burke-Ernzerhof exchange-correlation functional\cite{pbe}. The Brillouin-zone integration were performed using Monkhorst - Pack grids of special points with $16\times8\times4$ for total energy and structure optimizations and $32\times16\times8$ with tetrahedra method for electronic density of states and Fermi surface calculations. The spin-orbit (SO) interactions and the weak inter-layer van der Waals (vdW) interactions were all included in our calculations. We used grimme-d2\cite{grimme} vdW correction with parameter $london - s6 = 0.6$. The effect of electron correlations are included within DFT+U method with U = 3.0 eV for the Mo $4d$-states. Including electron-correlation brings the calculated band structure and Fermi-surface into excellent agreement with Quantum Oscillation and ARPES measurements as discussed in SI and also found in other very recent studies\cite{ldau1,ldau2}. Fermi-surface sheets and SdH orbits are visualized by our custom python code using Mayavi\cite{mayavi}. The quantum oscillation frequencies/orbits and their angle dependence were calculated using the skeaf code\cite{skeaf}.

\textbf{Crystal Synthesis.} Powder samples were prepared using the standard solid state synthesis method using high purity Mo powder (5N metals basis excluding W, Alpha Aesar), and Te shot (6N, Alpha Aesar). Large single crystals were grown using the Te self flux method as described in using the same source metals as for the powder samples. High sample quality has been confirmed by x-ray and neutron diffraction, stoichiometry has been confirmed by wavelength dispersive spectroscopy, and the samples measured have residual resistivity ratios greater than 1000.
\par
\textbf{Structural Measurements.} Determinations of the temperature and pressure dependent crystal structure were made using elastic neutron scattering measurements at 14.7 meV on the BT-4 triple axis spectrometer at the NIST Center for Neutron Research using a collimation and filter setup of open-pg-40'-pg-s-pg-40'-120' where pg refers to pyrolytic graphite. Single crystals were mounted in a steel measurement cell aligned in the H0L$_M$ zone and He was supplied as a pressure medium to maintain hydrostatic pressure conditions as described elsewhere \cite{Heikes2018}. The T$_d$ and T' phases and their volume fractions were identified from the position and intensity of (201)$_M$ reflections, which both split in 2$\theta$ and shift in $\omega$ in the T' phase. Rocking curves and $\omega$-2$\theta$ scans were taken at each pressure and temperature. Scans along (00L) from (2 0 0.5) to (2 0 4.5) were also obtained at 0.8 GPa in the all T$_d$ or T' condition, as well as in the mixed region region at both 0.8 and 1 GPa to look for possible superstructure reflections. None were observed. \par
\textbf{Transport Measurements.} A non-magnetic piston-cylinder pressure cell was used for transport measurements under pressure up to 1.8 GPa, choosing a 1 : 1 ratio of n-Pentane to 1-methyl-3-butanol as the pressure medium and superconducting temperature of lead as pressure gauge at base temperature. For transport measurements, we prepared a 110 $\mu m$ thick sample of MoTe$_{2}$ using four point contacts curing contacts with silver epoxy. Transport measurements up to 20 T were performed in Oxford Heliox insert and taken using in magnetic fields up to 14 T and down to 1.8 K in Physical Property Measurement System (PPMS). For superconducting temperature below 1.8 K, resistivity measurements down to 25 mK in a dilution refrigerator were taken using a Lakeshore LS370 AC resistance bridge. The resistivity values were taken by the average of 60 seconds stable and successive measurements. 
\section{Acknowledgements}

We thank J. D. Sau, Paul Kienzle and Daniel J. Campbell for helpful discussions.  The computational work was performed using the NIST Enki high-performance computer cluster. This work utilized facilities supported in part by the National Science Foundation under Agreement No. DMR-0454672 and NCNR/UMD Cooperative Agreement 70NANB15H261. Experimental measurements at the University of
Maryland were supported by DOE through Grant No. DE-SC-0019154, and materials synthesis by the Gordon and Betty Moore Foundation's  EPiQS Initiative through Grant No. GBMF4419. 

\section{Author contributions}
IL, CE, and TM performed transport measurements. CH and WR measured neutron scattering. NPB, CH, and SR synthesized materials. TY performed electron band structure calculations and theoretical analysis. IL, TY, CH, and NPB wrote the paper with contributions from all authors. 

\section{Competing Interests}
The authors declare no competing interests. 

\section{Materials   \&   Correspondence}
Correspondence and material requests should be directed to NPB.

\section{Data availability}
Data are available from the authors upon reasonable request.
\par


\end{document}


\beginsupplement
\title{Quantum oscillations from networked topological interfaces in a Weyl semimetal\\
Supplemental Material}
\author{
I-Lin Liu,$^{1,2,3}$
Colin Heikes,$^1$
Taner Yildirim,$^1$ 
Chris Eckberg,$^2$
Tristin Metz,$^2$
Sheng Ran,$^{1,2,3}$
William Ratcliff II,$^1$
Johnpierre Paglione,$^{2}$ 
and Nicholas P. Butch$^{1,2}$}
 
\affiliation{\vspace{4pt}
$^1$NIST center for Neutron Research, NIST, Gaithersburg, Maryland 20899, USA.
\\$^2$Center for Nanophysics and Advanced Materials, Department of Physics,  University of Maryland, College Park, MD 20742 USA.
\\$^3$Department of  Materials Science and Engineering, University of Maryland, College Park, MD 20742 USA.}

\date{\today}
\maketitle


\section{Elastic  neutron  scattering}
\begin{figure}
\centering
\subfigure{\label{Neutron}\includegraphics[scale=0.9, trim= 10 220 30 0,clip]{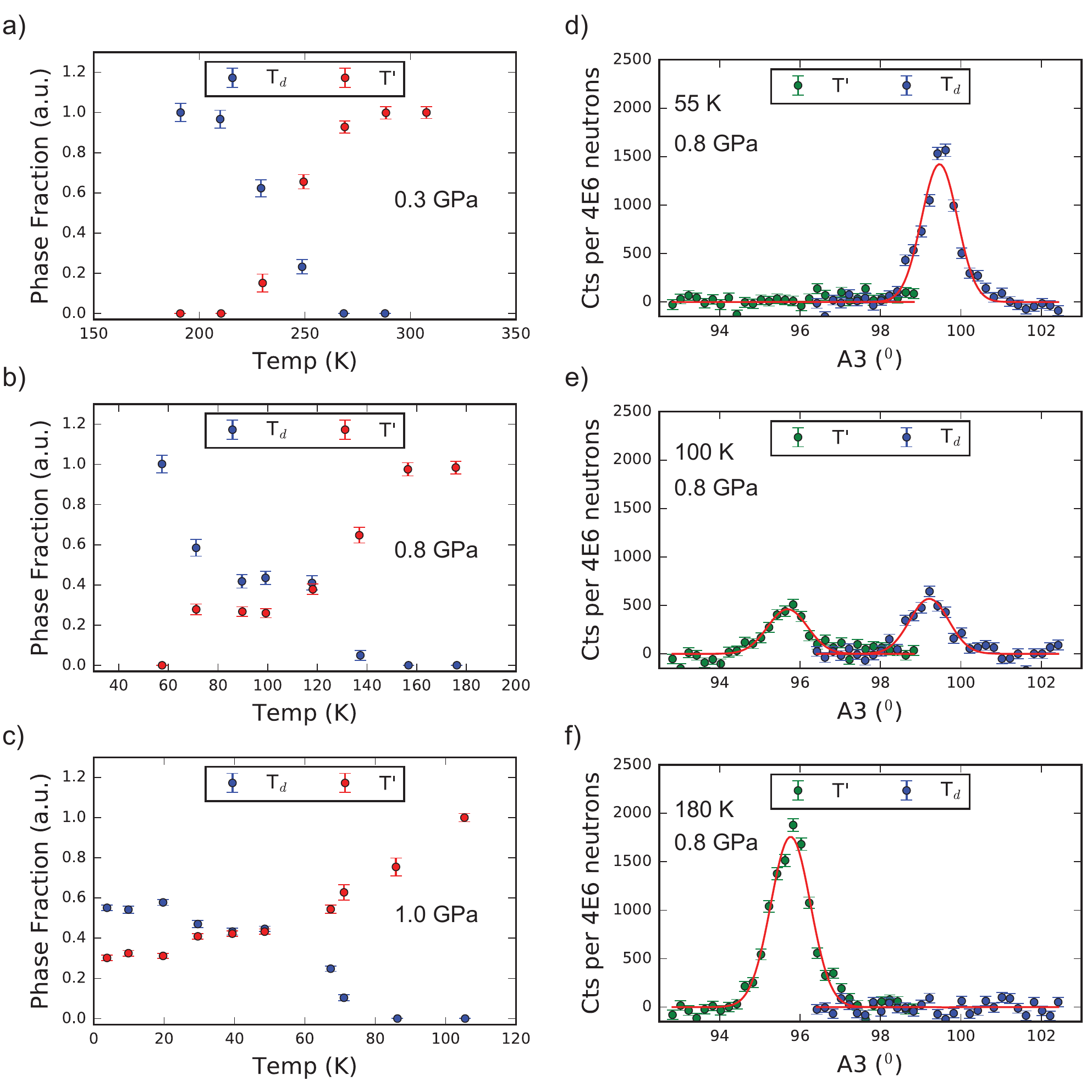}}
\subfigure{\label{Neutron-b}}
\subfigure{\label{Neutron-c}}
\subfigure{\label{Neutron-d}}
\subfigure{\label{Neutron-e}}
\subfigure{\label{Neutron-f}}
  \caption{Temperature evolution of T' and T$_d$ phase fractions. a-c shows the temperature dependent phase fractions of both the T' and T$_d$ phases at three different pressures illustrating the mixed region above 0.6 GPa. d-f show the background subtracted scattering intensities at the (201) peak positions in the pure T$_d$ phase, the mixed region, and in the pure T' phase used to extract the phase fractions for the 0.8 GPa data in panel (b). The red lines show the Gaussian fits to the data used to extract the integrated peak intensities for phase fraction determination. }
\end{figure}

\textbf{Structural phase determination and phase transition implications:}
The stated relationship between quantum oscillations and crystal structure is based on our pressure and temperature dependent elastic neutron scattering measurements. As is described in our previous work \cite{Heikes2018}, we determine the T$_d$ and T' phase fractions from the scattering intensity at the (201) reflections of the monoclinic unit cell for the T' phase. For the T$_d$ phase, this is for a unit cell with a $\beta$-angle of 90$^0$. The scattered intensity is determined from the integrated intensity of transverse scans centered at the nominal peak positions for each phase. We used the BT-4 triple-axis spectrometer at the NIST Center for Neutron research with a monochromatic 14.7 meV neutron beam and a collimation and filter geometry of Open-pg-40'-pg-s-pg-40'-120 where pg refers to an oriented pyrolytic graphite filter used to remove higher order neutrons. The temperature dependent integrated intensity of each phase, taken upon warming, is given in \cref{Neutron,Neutron-b,Neutron-c} at 0.3 GPa, 0.8 GPa, and 1 GPa respectively. \cref{Neutron-d,Neutron-e,Neutron-f} show the raw scattering data fit to a Gaussian peak shape used to extract the integrated intensity for the three different temperature regimes seen at 0.8 GPa. In the mixed region, the temperature dependent isobars show a clear deviation from normal mean field-like behavior expected for a first order phase transition. We attribute this to kinetic trapping of the structural phase transition. 

\par
\textbf{Kinetic freezing in a first order thermal structural phase transition:} Critical to our discussion of phase coexistence and our interpretation of the cause of the arrested transition is that despite an apparent suppression of a phase transition temperature to zero temperature at a specific pressure, we do not believe this should be described as a quantum structural phase transition. We attribute the effective "freezing" of the transition to the reduction of the structural transition temperature into a temperature regime where thermal fluctuations cannot overcome the energy barrier between the phases. We can draw an analogy between this behavior and the kinetic freezing observed in the low temperature first order magnetic transitions in (La,Pr,Ca)MnO$_3$ between a ferromagnetic and charge-ordered antiferromagnetic state \cite{Magglass2006_1}. In the magnetic case the complex energy landscape is attributed to competition between magnetic order, structural and chemical disorder, and inter-grain strain, which is consistent with other observations of similar transistions in magnetocaloric materials with magnetostrictive coupling \cite{Magglass2006_2}. Given the previously established sensitivity of the MoTe$_2$ structural phase transition to pressure and strain as well as the known structural disorder in van der Waals materials, it is reasonable to assume that inter-grain strain and disorder may dominate the phase transition kinetics in this system as well \cite{Heikes2018,yang2017strain}. At 0.8 GPa where we first observe a broad mixed region, the transition temperature is still high enough that it is possible to undercool the sample such that it eventually completes the phase transition, but at 1 GPa the transition temperature is further suppressed and no sufficient undercooling is possible leading to the mixed region as the ground state structure. 

\par
\begin{figure}
\centering
\subfigure{\includegraphics[scale=1, trim= 0 0 0 0,clip]{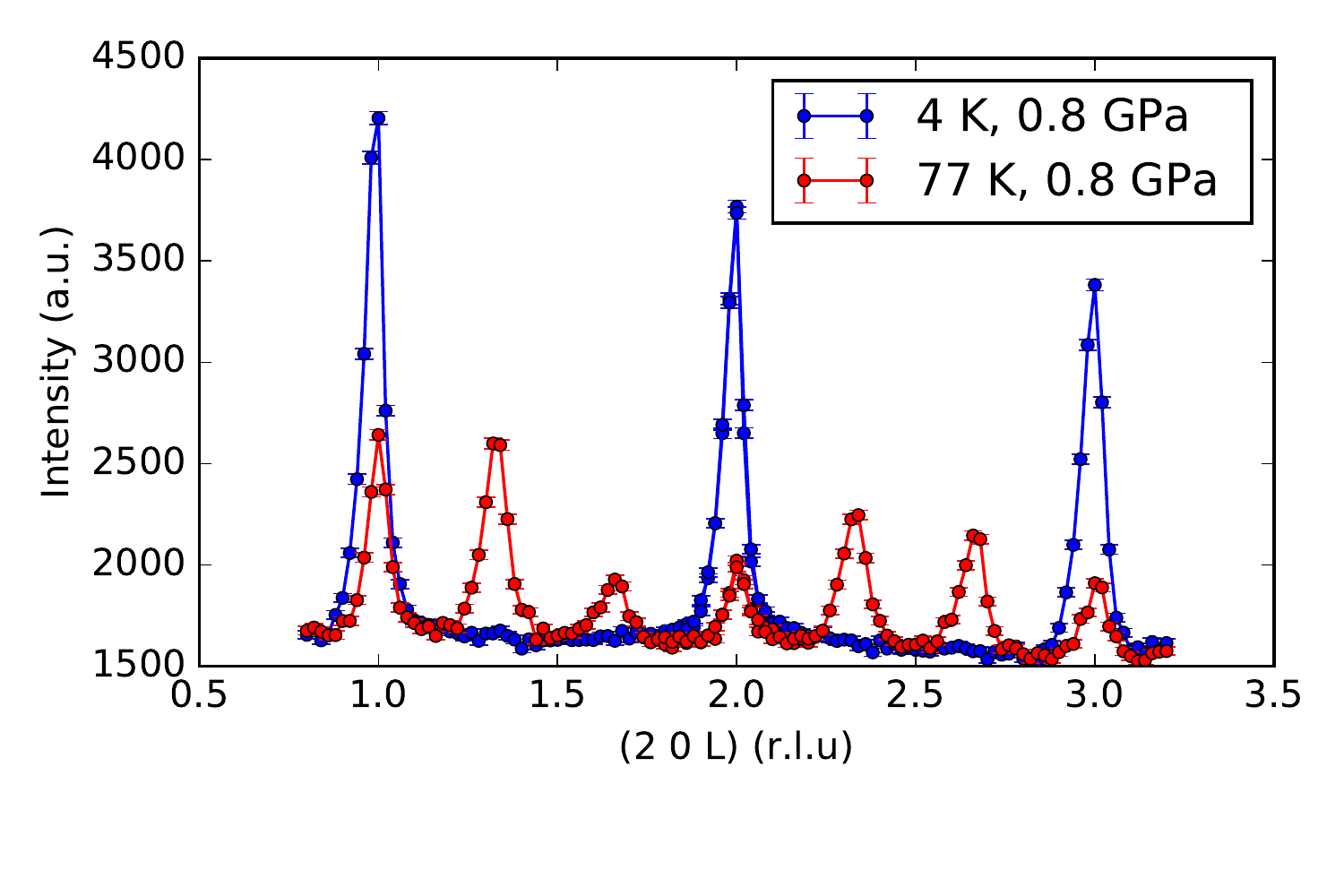}}
  \caption{\label{120KSI_Lscans}Scans along (20L) taken at 0.8 GPa showing scattering along L in the T$_d$ phase and in the mixed region. In the mixed region, we see no new scattering that would be indicative of stacking order along the c-axis, observing only clear reflections from both the T$_d$ and T' phases.  }
\end{figure}

\textbf{Structure in the mixed region:} Given the coherent oscillations in the mixed region, we considered whether this could be evidence of a new bulk structural phase driven by stacking disorder during the phase transition. If this were the case, we might expect to observe new reflections along the stacking direction. We performed [00L] scans along [20L] at 0.8 GPa at temperatures corresponding to the T$_d$ phase and the mixed region. These are shown in figure S1a. Clearly we see no evidence of new reflections that would indicate a new phase, instead observing only integer L peaks for the T$_d$ phase, split integer peaks from both monoclinic twins in the T' phase, and scattering attributable to the mixed phase in the mixed region. While we cannot completely rule out a new intermediate meta-stable phase, we see no indication of this either from superlattice reflections along [00L] or from missing intensity in the mixed region. The same is seen at 1.5 K and 1 GPa, with a slightly different ratio of T$_d$ to T' phase, consistent with the volume fractions extracted from the (201) scans used for the phase diagram in figure ~\ref{120KSI_Lscans}.  

\clearpage
\section{First-Principles Calculations}

 In this section, we discuss the details of our first-principles electronic structure calculations of MoTe$_2$ as a function of pressure for both $1T'$-   and $T_d$-phases, respectively. 

\subsection{Method}
The total energy, structure optimization under pressure, and band structure and Fermi surface calculations were performed by  Quantum Espresso,\cite{qe},   which is based on density-functional theory (DFT), using a plane wave basis set and  fully relativistic  all-electron projected augmented wave  (PAW) potentials\cite{paw1,paw2}.  The $4s, 4p, 4d$, and $5s$ electrons of Mo and the 4d, 5s, and 5p electrons of Te were treated as valence. We used 0.02 Ry Methfessel-Paxton smearing with wavefunction and charge density cut-off energies of 100 Ry and 800 Ry, respectively.  The exchange-correlation interactions were  described by the generalized gradient approximation (GGA) with  the Perdew-Burke-Ernzerhof exchange-correlation functional\cite{pbe}. The Brillouin-zone integration were performed using Monkhorst-Pack grids of special points with $16\times8\times4$ for total energy and structure optimizations and $32\times16\times8$ with tetrahedra method for electronic density of states and Fermi surface calculations.   The spin-orbit (SO) interactions and the  weak inter-layer van der Waals (vdW) interactions were all included in our calculations.  We used grimme-d2\cite{grimme} vdW correction with parameter $london-s6=0.6$. The effect of electron correlations are included within DFT+U method with U=3.0 eV for the Mo $4d$-states.  Including electron-correlation brings the calculated band structure and Fermi-surface into excellent agreement  with Quantum Oscillation and ARPES measurements as found in  very recent studies\cite{ldau1,ldau2}. Fermi-surface sheets and SdH orbits are visualized by our custom python  code  using  Mayavi\cite{mayavi}. The quantum oscillation frequencies/orbits and their angle dependence were calculated using the skeaf code\cite{skeaf}.
\subsection{Effect of Electron Correlations on the Band Structure and Fermi Surface of MoTe$_2$}
Recent studies\cite{ldau1,ldau2} found that electron correlations are essential for a precise description of the bulk electronic structure of T$_d$-MoTe$_2$ as revealed by angular resolved photoemission spectroscopy (ARPES)\cite{ldau1} and the angular dependence of the Fermi surface by quantum oscillation (QO) experiments\cite{ldau2}. Hence, in our study we adopted DFT+U scheme to describe the electron correlations within the Mo $4d$-states. The overall best agreement with ARPES and QO data is obtained for U=3 eV\cite{ldau1,ldau2}, which was also used in our calculations in this study.

\begin{figure}[b]
\includegraphics[scale=0.6]{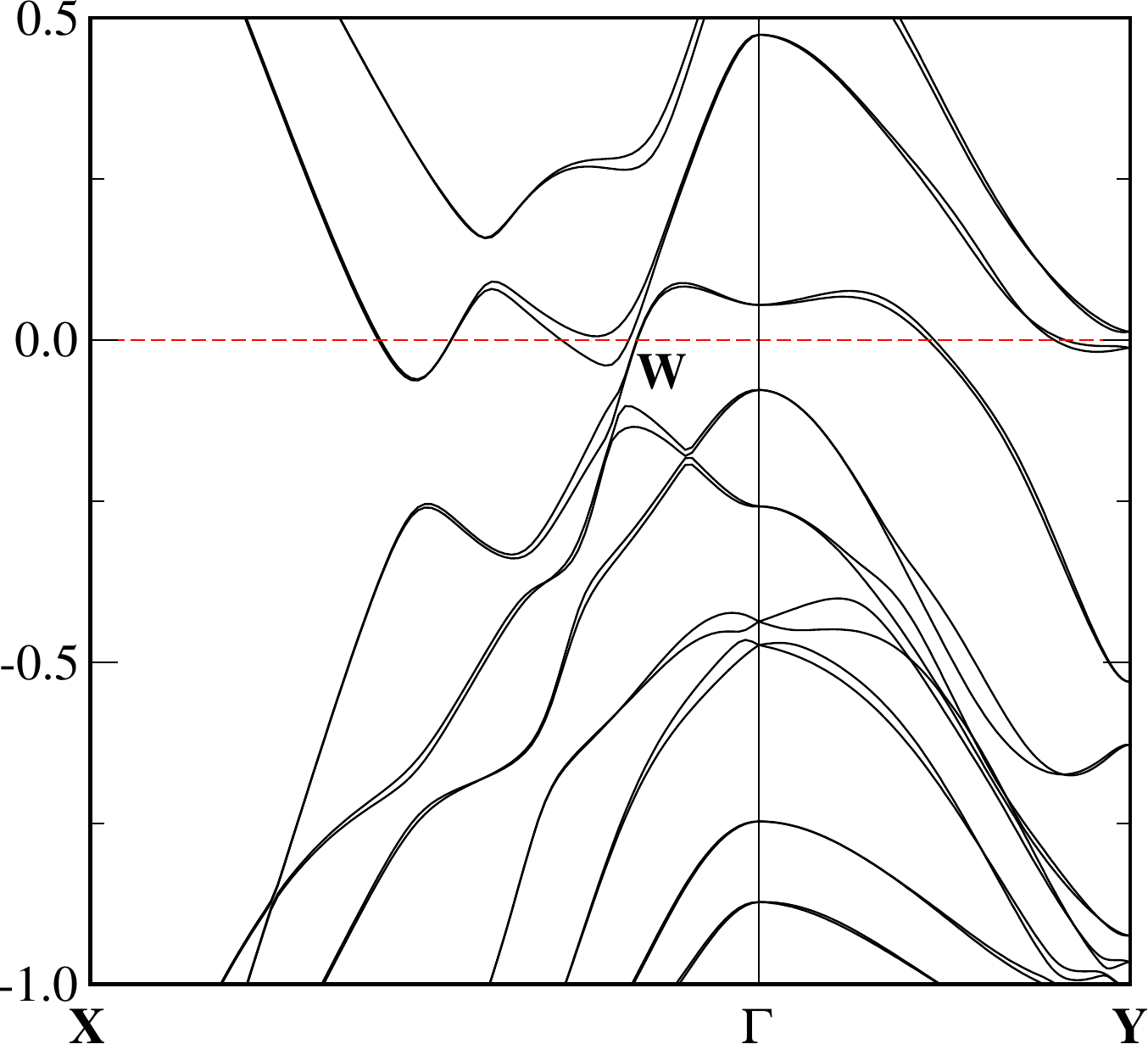}
\includegraphics[scale=0.6]{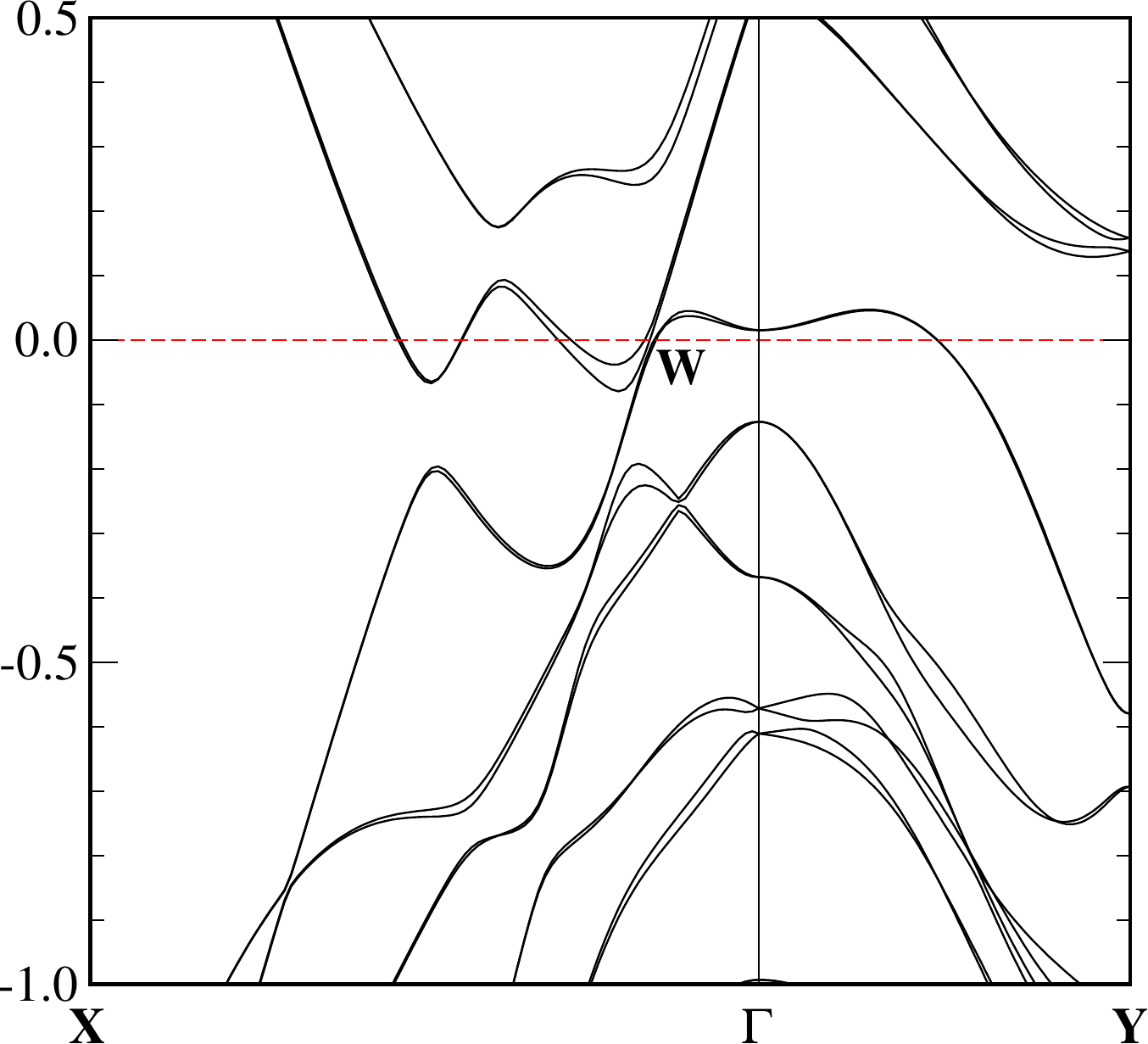}  
\caption{\label{bandstr_Td}  Electronic band structure without (left) and 
with Hubbard $U$ (right), showing the effect of the electron correlations on
the band structure. The main effect is to shift up the bands near Fermi
level around Y-point while the Weyl point (shown as "w") was not effected
by $U$. The small splitting of the bands are due to Spin-Orbit (SO) coupling
and the lack of inversion symmetry in the T$_d$ phase.
}
\end{figure}
\begin{figure}
\includegraphics[scale=0.6]{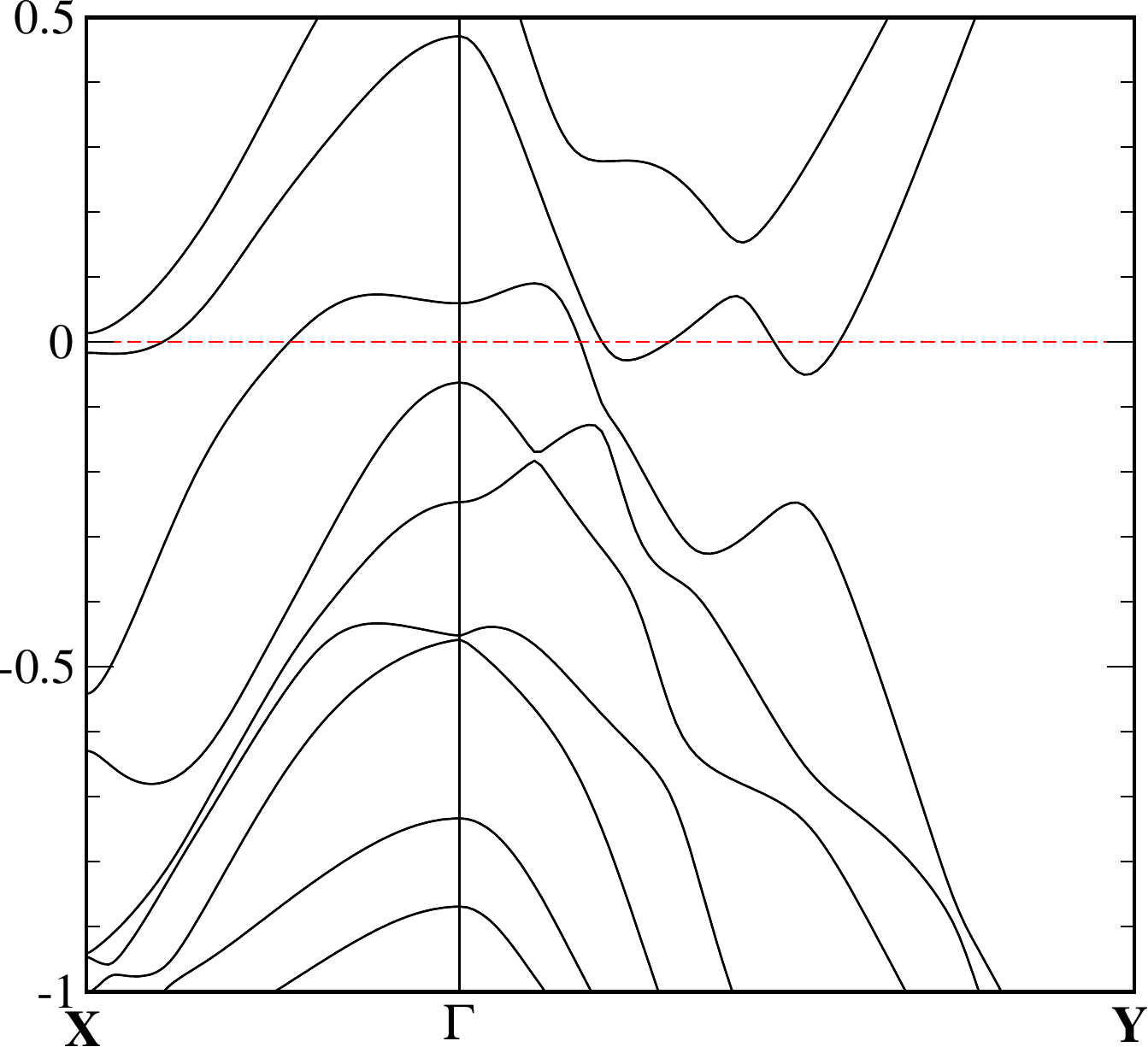}
\includegraphics[scale=0.6]{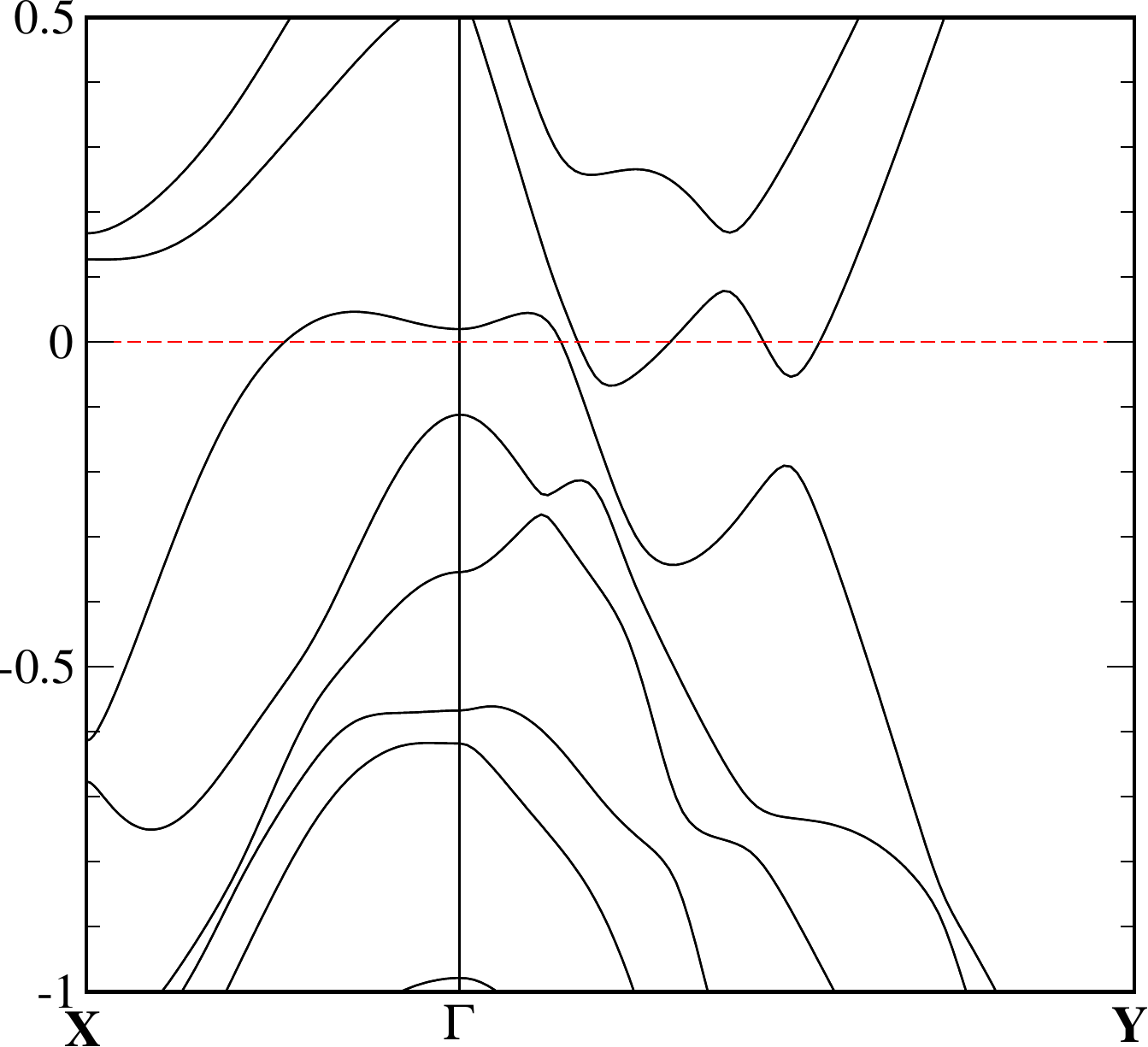}  
\caption{\label{bandstr_Tp}  
Band structure with (right) and without
$U$ term (left) in $1T'$-phase of bulk MoTe$_2$.  The effect of SO coupling is
still important even though it does not split the bands (but shift them
around to effect the Fermi Surface).
}
\end{figure}

\begin{figure}
\includegraphics[scale=0.7,trim= 0 100 0 0,clip ]{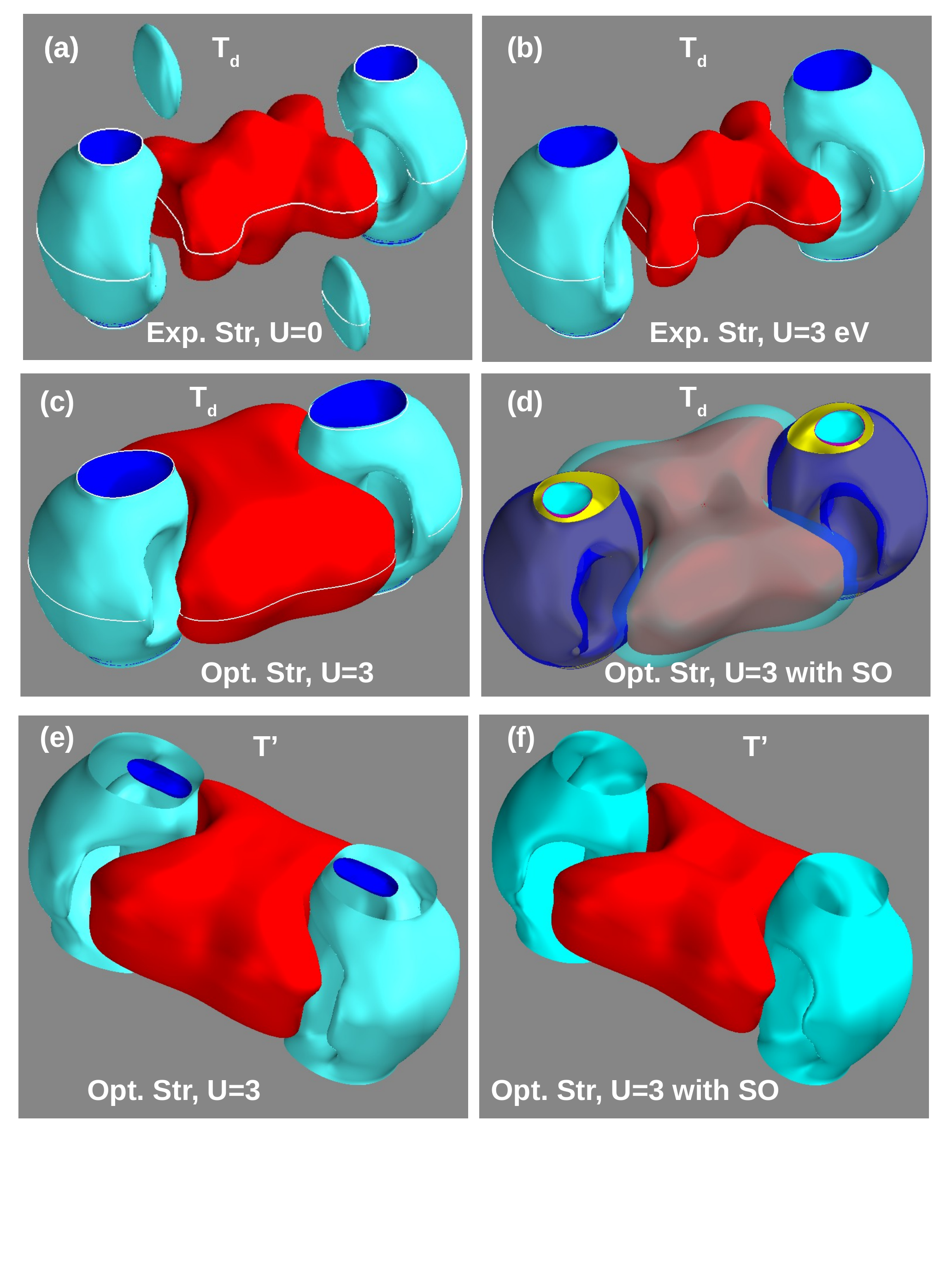}
\caption{\label{varFS}  Fermi surface plots of MoTe$_2$ for various cases;
(a) Experimental structure and without U term; (b) Experimental structure with
U=3 eV; main effect of which is to remove the states near Y-point;
(c) Fully optimized structure with U=3 eV; Note that it is quite different
than one from experimental structure shown in b; (d) Fully optimized structure with
U=3 eV and also with spin-orbit (SO) coupling; The main effect of spin-orbit coupling
is to shrink and expand the surfaces so that they  split; Note the significant
shrinkage of the surface shown as dark blue color; (e-f) shows the Fermi surface in
the $1T'$ phase with optimized structure and U=3 eV.   Due to inversion symmetry, t
here is no splitting of the  Fermi surface in $1T'$-phase 
but the bands are shifted around and the resulting Fermi surface
is different. Also note that the small electron packet (dark blue in (e))   is removed 
with the inclusion of SO coupling (f).  
}
\end{figure}

Figures~\ref{bandstr_Td} and ~\ref{bandstr_Tp} show the effect of the Hubbard U (taken as 3.0 eV) on the band structure and the Weyl-points in the $T_d$- and $1T'$-phases of MoTe$_2$, respectively.  
 We note that the biggest effect is to shift the bands near Y-point so that we do have any electron pocket at the Fermi surface as shown in Fig. \ref{varFS}. 
The other main effect is to lower some of the bands  further below Fermi level which does not have any effect on the Fermi surface.  It is important to note that the Weyl point near Fermi level remains unaffected with the inclusion of the Hubbard correlation term $U$ in our calculations. However, as we shall see below, including U is critical to explain the pressure dependence of the QO frequencies that we have measured in this study.
\begin{table}
\caption{\label{exp_cal_Td}
Lattice parameters and fractional 
atomic positions as determined by our neutron diffraction measurements and
our DFT+U calculations for $T_d$  MoTe$_2$ (\textit{Pmn2$_1$}).     }
	\begin{ruledtabular}
		\begin{tabular}{lcc}
Experimental Structure   &  	& 	 	 \\
a=3.46464  \AA, 	b=6.30716 \AA	&      	c=13.84310  \AA    &  90$^o$ 90$^o$ 90$^o$      \\
			\hline
Mo \hfill    0.000000000     &    0.606100004      &   0.497243989 \\
Mo \hfill    0.500000000    &     0.393899996    &     0.997243989 \\
Mo  \hfill   0.000000000   &      0.029300001    &     0.014242000 \\
Mo \hfill    0.500000000   &      0.970700018    &     0.514242010 \\
Te \hfill    0.000000000   &      0.865899961    &     0.653545972 \\
Te \hfill    0.500000000   &      0.134100020    &     0.153545972 \\
Te \hfill    0.000000000   &      0.641099962    &     0.112019999 \\
Te \hfill    0.500000000   &      0.358900000    &    0.612019999 \\
Te  \hfill   0.000000000  &       0.287699989    &     0.857258999 \\  
Te  \hfill   0.500000000   &      0.712299993   &      0.357259033 \\
Te   \hfill   0.000000000  &       0.214699994  &       0.401510016 \\
Te  \hfill    0.500000000  &       0.785300043  &       0.901509982 \\ \hline
DFT+U Optimized Structure   &  	& 	 	 \\
a=3.51242  \AA, 	b=6.33797 \AA	&      	c=13.80214  \AA    &  90$^o$ 90$^o$ 90$^o$      \\
			\hline
 Mo    \hfill   0.000000000  & 0.596570039 &  0.499028659 \\
Mo   \hfill     0.500000000 &  0.403429961 &  0.999028659 \\
Mo  \hfill      0.000000000 &  0.043489108 &  0.012956972 \\
Mo  \hfill      0.500000000 &  0.956510911 &  0.512956982 \\
Te  \hfill      0.000000000 &  0.854056153 &  0.653752430 \\
Te  \hfill      0.500000000 &  0.145943828 &  0.153752430 \\
Te  \hfill      0.000000000  & 0.649980885 &  0.109058347 \\
Te  \hfill      0.500000000 &  0.350019077 &  0.609058347 \\
Te  \hfill      0.000000000 &  0.302836245 &  0.858321051 \\
Te  \hfill      0.500000000 &  0.697163737 &  0.358321085 \\
Te \hfill       0.000000000 &  0.203297619 &  0.402703515 \\
Te  \hfill      0.500000000 &  0.796702418 &  0.902703481 \\ 
		\end{tabular}
	\end{ruledtabular}
\end{table}

\begin{table}
\caption{\label{exp_cal_Tp}
Lattice parameters and fractional 
atomic positions as determined by our neutron diffraction measurements at 300 K
and our DFT+U calculations for $T'$  MoTe$_2$ (\textit{P2$_1$/m}).     }
	\begin{ruledtabular}
		\begin{tabular}{lcc}
Experimental Structure   &  	& 	 	 \\
a=6.3281  \AA, 	b= 3.4770 \AA	&      	c=13.021  \AA    &  90$^o$ 93.882$^o$ 90$^o$      \\
			\hline
Mo \hfill    0.182799990  &       0.250000000  &       0.008300000 \\
Mo \hfill    0.817199966  &       0.750000017  &       0.991700002 \\
Mo \hfill    0.319399986  &       0.750000017  &       0.506199997 \\
Mo \hfill    0.680599966  &       0.250000000  &       0.493799979 \\
Te \hfill    0.587999989  &       0.250000000  &       0.106399996 \\
Te \hfill     0.411999994  &       0.750000017  &       0.893599938 \\
Te \hfill    0.096600004  &       0.750000017  &       0.149299988 \\
Te \hfill    0.903399993  &       0.250000000  &       0.850699971 \\
Te \hfill    0.557100023  &       0.750000017  &       0.351300002 \\
Te \hfill    0.442900004  &       0.250000000  &       0.648699974 \\
Te \hfill    0.056299998  &       0.250000000  &       0.395299983 \\
Te \hfill    0.943700035  &       0.750000017  &       0.604699925 \\ \hline
DFT+U Optimized Structure   &  	& 	 	 \\
a=6.3422  \AA, 	b=3.5106 \AA	&      	c=13.8292  \AA    &  90$^o$ 93.8907$^o$ 90$^o$      \\
			\hline
Mo \hfill      0.181248302  & 0.250000000 &   0.007402050 \\
Mo \hfill      0.818751654  & 0.750000017 &  0.992597952 \\
Mo \hfill      0.320793055  & 0.750000017 &  0.506405748 \\
Mo \hfill      0.679206897  & 0.250000000 &  0.493594228 \\
Te \hfill      0.589309059  & 0.250000000 &  0.103197530 \\
Te \hfill      0.410690924  & 0.750000017 &  0.896802404 \\
Te \hfill      0.097660263  & 0.750000017 &  0.147835158 \\
Te \hfill      0.902339734  & 0.250000000 &  0.852164801 \\
Te \hfill      0.559332448  & 0.750000017 &  0.352629203 \\ 
Te \hfill      0.440667579  & 0.250000000 &  0.647370773 \\
Te \hfill      0.057121356  & 0.250000000 &  0.396415521 \\
Te \hfill      0.942878677  & 0.750000017 &  0.603584387 \\
		\end{tabular}
	\end{ruledtabular}
\end{table}

 In our calculations, besides the Hubbard term $U$, there are other parameters such as lattice constants and atomic positions that we need to determine. One way is to use experimental parameters or to determine them self consistently within the DFT+U structural optimization at any given pressure. Figure\ref{varFS} shows the Fermi surface of $T_d$- and $1T'$ phases of MoTe$_2$ for both experimental and DFT optimized structures with and without electron correlation effects (i.e. U). We note that the Fermi surface is very sensitive to the lattice parameters
 and the atomic positions. Our optimized lattice parameters and atomic positions are within $1\%$ the experimental values  as shown in Tables~\ref{exp_cal_Td} and~\ref{exp_cal_Tp}. The biggest variation is in the a-axis for 
 the $T_d$ phase and it's $1.4\%$. Despite this excellent agreement between experiment and calculations, the difference in the Fermi surface between experimental and the optimized structures is quite large.  
 In order to be self-consistent, we decided
 to use optimized lattice parameters and atomic positions for a given pressure as obtained from our DFT+U calculations. In this way, we are able to determine the pressure dependence of the Fermi Surface and determine
 the quantum oscillation orbits and frequencies.
  The  only free-parameter in our DFT+U calculations is the Hubbard U, which was shown to be around U=3 eV for MoTe$_2$ to match the ARPES measurements as well as the angle dependence of the QO  frequencies\cite{ldau1,ldau2}.

\subsection{Pressure Dependence of the Fermi Surface and Quantum Oscillations}
 
\begin{figure}
\includegraphics[scale=0.7,trim= 0 40 100 0,clip ]{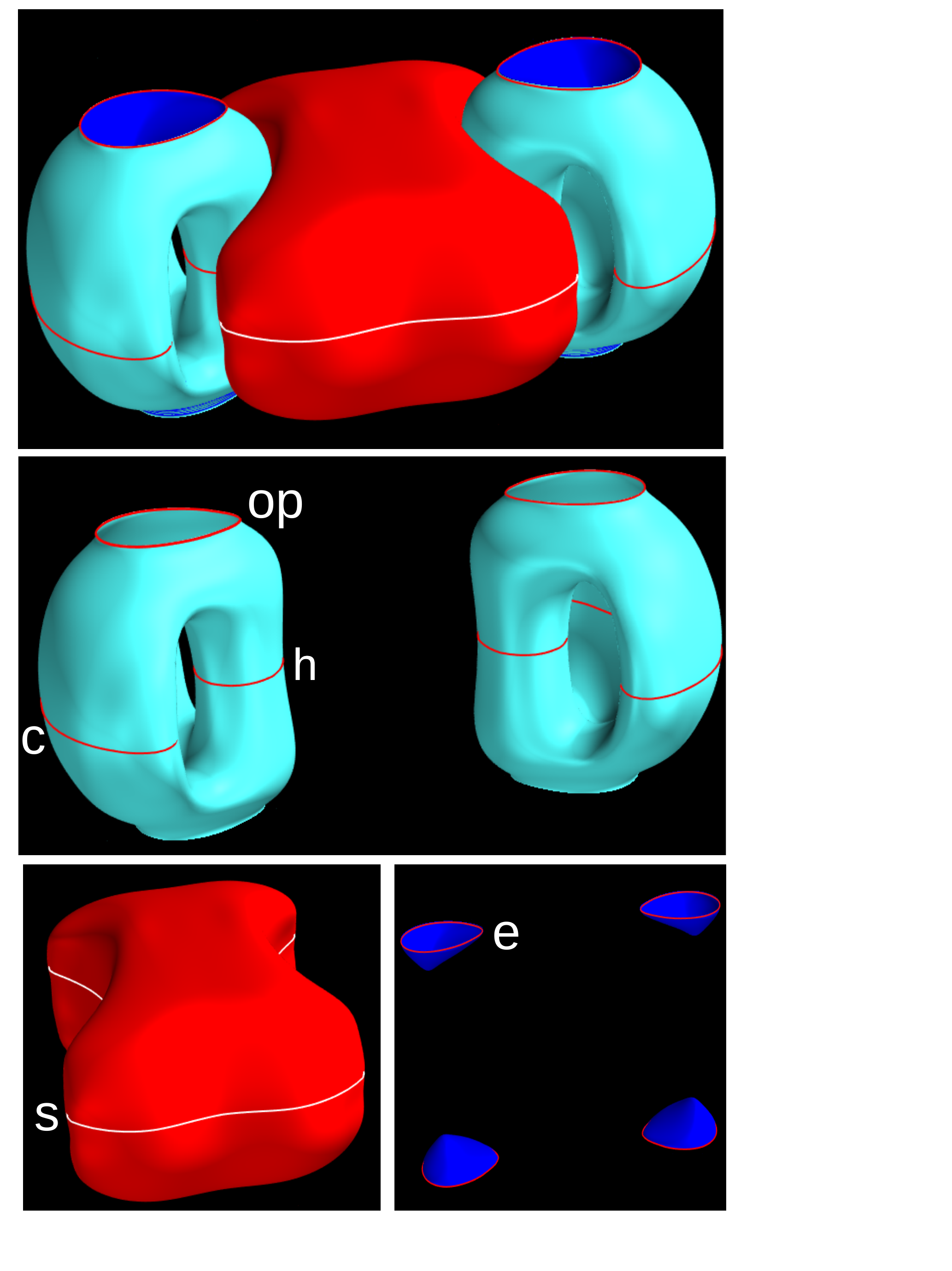}
\caption{\label{FSorbits}  Top panel shows the main shape of the Fermi surface
in both phase of MoTe$_2$ which consists of three types of bands; The red
one is a {\bf s}quare like box shape with mainly hole-character. We denote the orbits
around this surface as "{\bf s}". The main portion of the Fermi surface is electron-like
and has the coffee mug shape (light blue). This shape supports three possible extreme
orbits as shown as "{\bf op}" (which is at the {\bf op}ening of the surface), as "{\bf h}" which 
is the orbit around the {\bf h}andle like shape, and "{\bf c}" which is the orbit around
the {\bf c}up portion of the Coffee Mug-shape Fermi surface. Interestingly, this 
cup-like shape is the most sensitive to the pressure  and
the frequency of this orbit increases with increasing pressure, in excellent
agreement with our measurements. 
Finally, we have small pocket of electron Fermi surface (dark blue), which has
the {\bf e}llipsoidal shape and therefore denoted as "{\bf e}".
}
\end{figure}

In this section, we present our results related to the pressure dependence of the
Fermi surface and quantum oscillation orbits as a function of applied pressure
for both phases of MoTe$_2$.  In both phases, we have similar Fermi surface and
orbits which are summarized in Figure~\ref{FSorbits}. Near $\Gamma$, we have a square-box like
Fermi surface (red). The orbit around this Fermi surface is shown in Fig.\ref{FSorbits}
as "s".  Then, we have an electron-like Fermi surface with a shape of a coffee mug (light
blue). This shape has basically three types of extremal orbits as shown in the
Figure~\ref{FSorbits}. We label the orbits at the opening as "op". Then, the orbit
near the handle like surface as "h". And, finally we have the orbits near the cup like
denoted as "c". As we shall see below, this orbit is sensitive to pressure and we
identified this orbit as in our quantum oscillation measurements.  Finally, we have
small pocket of squashed elliptical surface (dark blue), which we call it "e" orbit.
In the case of $T_d$-phase, these orbits have two counterparts due to SO-splitting.

 We have carried out full structure optimization at a given pressure and then 
 calculate the Fermi surface over a dense k-grid to determine the orbit frequencies 
 using skeaf code\cite{skeaf}. Our results are summarized 
 in Tables~\ref{orbits_Td} and~\ref{orbits_Tp}. We note that most
of the orbit-frequencies do not change much with applied pressure but the
cup-orbit increases with increasing pressure. As discussed in the
main text, the slope of the frequency increase with pressure is in excellent
agreement with the observed shifts in the experimental measurements.
Hence, we identified this orbit as the one probed in our quantum oscillation 
measurements.

\begin{table*}[ht]
\caption{\label{orbits_Td} 
QO Frequencies (kT) in $T_d$-MoTe$_2$ phase as a function of pressure (kbar).
The orbit labels are defined in Figure~\ref{FSorbits}. }
\centering
\begin{tabular}{
p{0.06\linewidth}p{0.06\linewidth}p{0.06\linewidth}
p{0.06\linewidth}p{0.06\linewidth}p{0.06\linewidth}
p{0.06\linewidth}p{0.06\linewidth}p{0.06\linewidth}
p{0.06\linewidth}p{0.06\linewidth}p{0.08\linewidth}
}  
\hline
Orbits  & 0 kbar  & 2 kbar & 4 kbar &
6 kbar & 8 kbar & 10 kbar &
12 kbar & 14 kbar & 16 kbar &
18 kbar & 20 kbar  
\\
\hline
 h1 & 0.14 & 0.15 &  0.15 & 0.155 &  0.16& 0.163   &  0.167  & 0.17 &  0.174 & 0.17 & 0.17 \\
 h2 & 0.1 & 0.10 & 0.10 & 0.11 & 0.11& 0.104& 0.11 & 0.12 & 0.116 & 0.12 & 0.18 \\ \hline
 op1 & 0.235 & 0.23 & 0.21 & 0.22& 0.21& 0.22& 0.21 & 0.21 & 0.210 & 0.204 & 0.19 \\
 op2 & 0.22 & 0.22 & 0.21 & 0.207& 0.20& 0.20 & 0.19 & 0.19 & 0.196 & 0.186 & 0.18 \\ \hline
 c1 & 0.33 & 0.37 & 0.405 & 0.44 & 0.470 & 0.506 & 0.57 & 0.57 & 0.601 & 0.653 & 0.684 \\
 c2 & 0.294 & 0.33 & 0.36& 0.395& 0.426& 0.455 & 0.51 & 0.51 & 0.542 & 0.600& 0.626 \\ \hline
 e1 & 0.08 & 0.06 & 0.06 & 0.055& 0.06& 0.05 & 0.046 & 0.045 & 0.05 & 0.056 & 0.04 \\
 e1 & 0.08 & 0.08 & 0.08 & 0.07& 0.07& 0.07 & 0.06 & 0.06 & 0.06 & 0.056 & 0.06 \\ \hline
 s1 & 1.90 & 1.9 & 1.98 & 2.01 & 2.04& 2.07 & 2.10 & 2.13 & 2.15 & 2.21 & 0.26-2.24 \\
 s2 & 2.16 &2.2 & 2.24 & 2.27 &2.30  & 2.33& 2.36 & 2.39 & 2.41 & 2.42 & 0.29-2.5 \\
\hline
\end{tabular}
\end{table*}

\begin{table*}[ht]
\caption{\label{orbits_Tp} QO Frequencies (kT) in $T'$-MoTe$_2$ phase as a function of pressure (kbar). 
The orbit labels are defined in Figure~\ref{FSorbits}. 
}
\centering
\begin{tabular}{
p{0.06\linewidth}p{0.06\linewidth}p{0.06\linewidth}
p{0.06\linewidth}p{0.06\linewidth}p{0.06\linewidth}
p{0.06\linewidth}p{0.06\linewidth}p{0.06\linewidth}
p{0.08\linewidth}p{0.08\linewidth}p{0.08\linewidth}
}  
\hline
Orbits  & 0 kbar  & 2 kbar & 4 kbar &
6 kbar & 8 kbar & 10 kbar &
12 kbar & 14 kbar & 16 kbar &
18 kbar & 20 kbar  
\\
\hline
 h & 0.14 & 0.132   & 0.136 & 0.14 & 0.144 & 0.147 & 0.150 & 0.151 &  0.152 & 0.154 &  0.153 \\
 op & 0.33 & 0.29   & 0.293 & 0.291& 0.294 & 0.294 & 0.295 & 0.291 &  0.291 & 0.292 & 0.292 \\  
 c & 0.26 & 0.36   & 0.392 & 0.425& 0.457 & 0.489 & 0.520 & 0.556 &  0.5854 & 0.616 & 0.646 \\
 s & 1.896 & 2.027 & 2.030 & 2.10 & 2.130 & 2.160 & 2.185 & 2.245 & 0.24-2.25 & 0.47-2.27 & 0.63-2.3 \\  
\hline
\end{tabular}
\end{table*}

\begin{figure}
\includegraphics[scale=0.6, trim= 0 80  0 0]{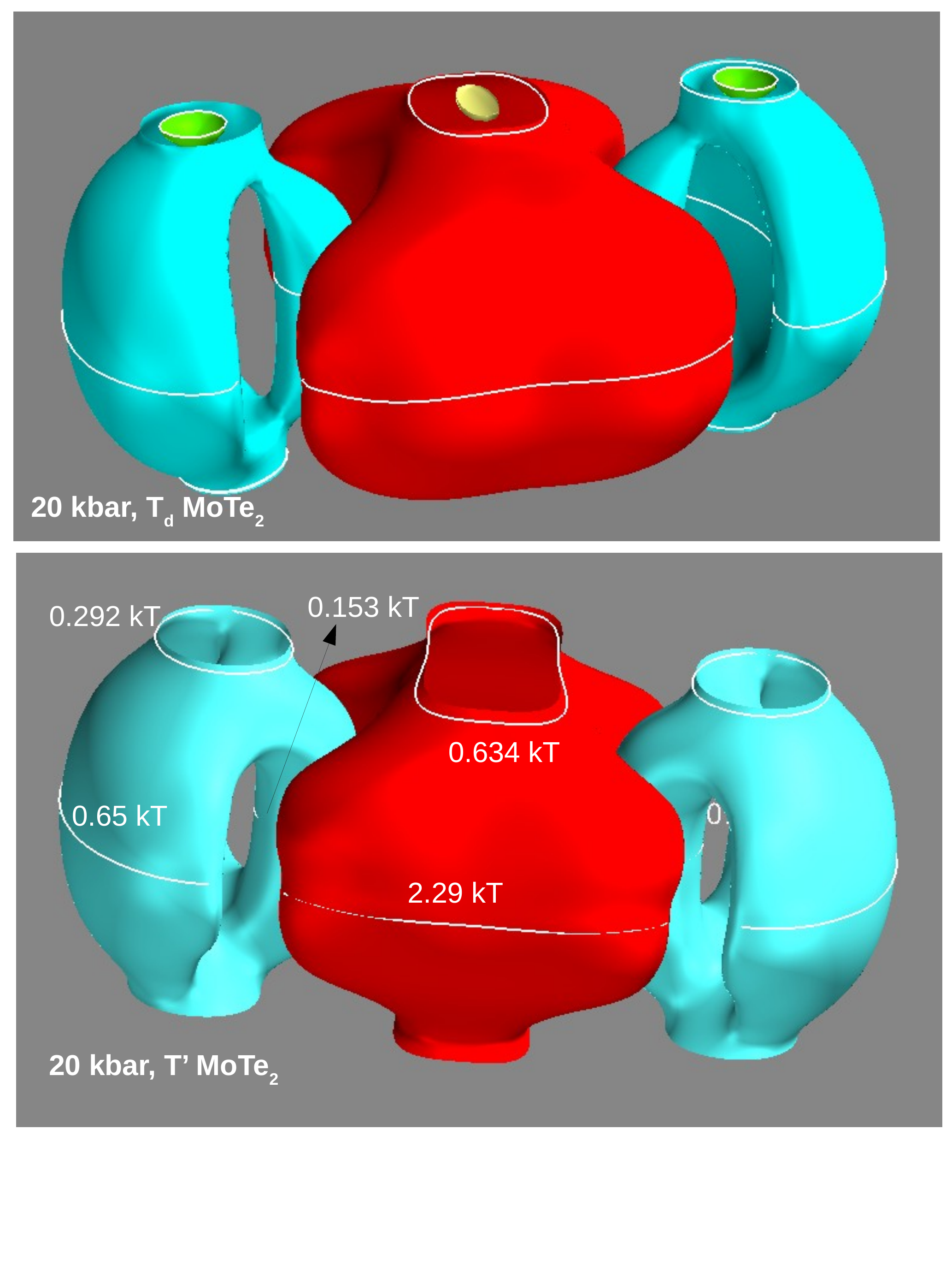}
\caption{\label{FS20kbar}  Fermi surface of MoTe$_2$ at 20 kbar
pressure for $T_d$ phase (top) and for $1T'$-phase (bottom).
Note that the center square-box like Fermi surface start to have
opening at the top with orbit frequencies near 0.65 kT, which is
similar to the cup-site orbit.
}
\end{figure}

The topology of the Fermi surface pretty much stays the same with applied pressure
up to 16 kbar (i.e. 1.6 GPa). Due to smaller lattice constants the band overlap 
gets larger with increasing pressure which in turn increases the orbit frequencies.
However at pressures larger than 1.6 GPa, due to strong inter-layer interaction, the
hole-band with the square box shape starts to have an opening at the top/bottom
of the box-surface as shown in Figure~\ref{FS20kbar}. For comparison, we show
Fermi Surface at 20 kbar for both phases in Fig.~\ref{FS20kbar} but we note
that at these pressures, the main phase is the $1T'$ phase where we have
inversion symmetry. Interestingly the new orbit at the top of the hole-square band
has about the same oscillation frequency as the cup-orbit near 0.65 kT.

\clearpage

\subsection{Angle dependence of Shubnikov-de Haas oscillations}
\begin{figure*}[ht]
\centering
 \subfigure{\label{FA}\includegraphics[scale=0.61, trim= 3 82 10 75,clip]{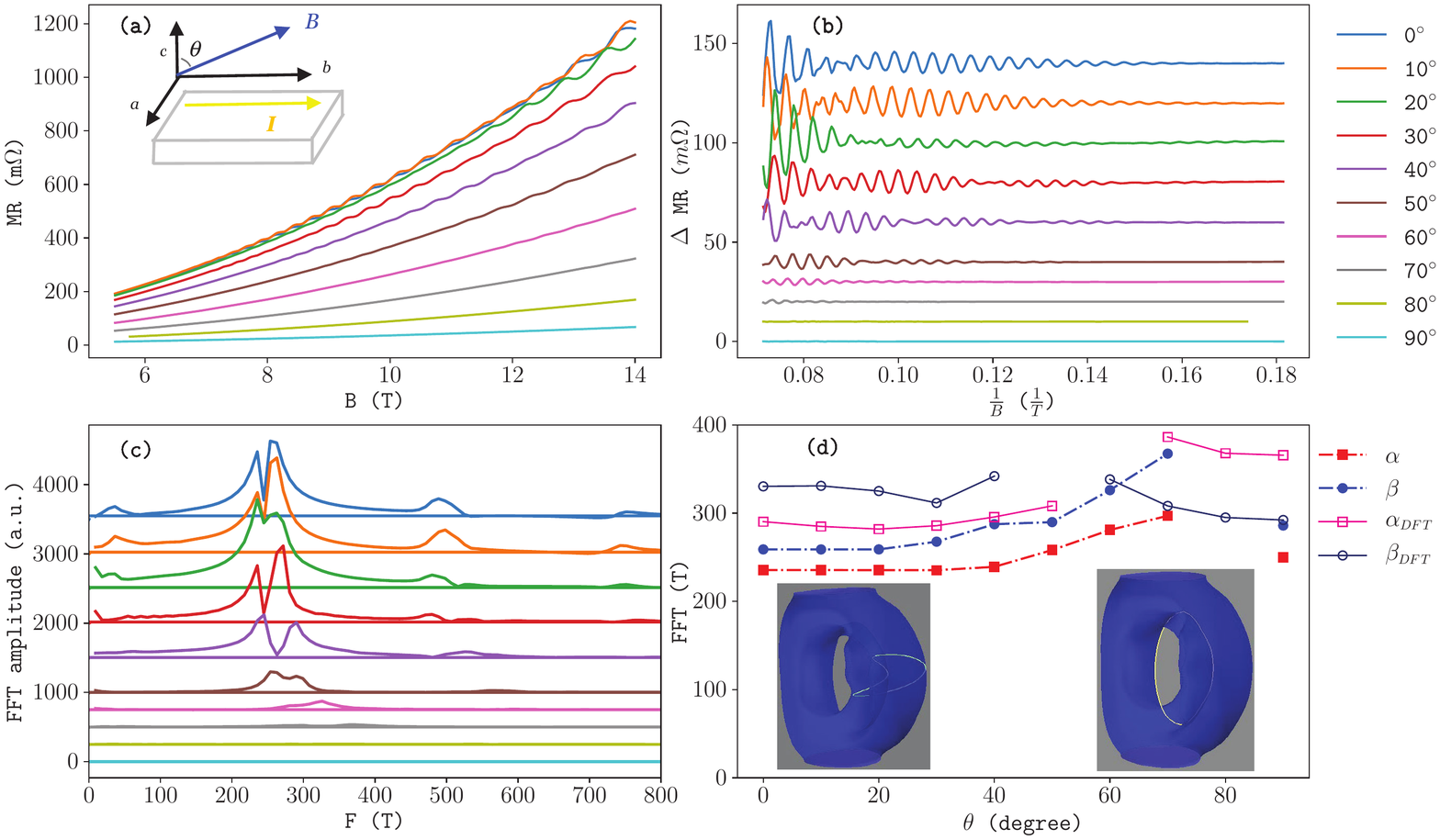}}
 \subfigure{\label{FA-b}}
  \subfigure{\label{FA-c}}
   \subfigure{\label{FA-d}}
  \caption{(a) The angular longitudinal MR of the bulk MoTe$_2$ sample measured at ambient pressure and 1.8 K when magnetic field moves from $c$ axis to $b$ axes ( $\vec{B}$ // $c$ axis, ${\theta}$= 0). Schematic represents the four-point magnetoresistance measurement, which current flows along $b$ axis, magnetic field moves along $c$ axis to $b$ axis with angle $\theta$. (b) The corresponding SdH oscillations were observed by $\Delta\rho=\rho-\rho_{background}$ above 5.5 T to 14 T. (c) The FFT spectrum of SdH oscillation as a function of the frequency show three pockets, whose oscillation frequencies slightly increase above ${\theta}$ > $50^{\circ}$.  (d) The extracted angle-dependence of the oscillation frequencies along with DFT-calculations. The angle dependence of the quantum oscillations for the
cup-shape orbit as the magnetic field is turned from c-axis (left) to b-axis (right).
The insets show the Fermi surface and the corresponding orbits; For field direction
near 40-60$^{\circ}$, due to two closely related Fermi surfaces due to SO interactions
and complicated MUG-shape surface, the extremal orbit is rather complicated involving both
the cup-portion of the surface and the interior of the MUG-shape surface, which may
explain the weak intensity of the oscillations for field directions near b-axis.
  }
\end{figure*}

{\label{AQO}}
Here we show angular magnetoresistance measured at ambient pressure and 1.8 K when magnetic field rotates from $c$ axis to $b$ axis as the insert figure in Fig.~\ref{FA}. The beat frequency increases with the increasing of ${\theta}$, the angle between magnetic field and $c$ axis as shown in Fig.~\ref{FA-b}. The cross section areas of two electron bands, $F_{\alpha}$ and $F_{\beta}$ slightly increase with increasing of ${\theta}$ at 1.8 K. 

In order to understand this weak angle dependence of the SdH oscillations, we have calculated the
extreme orbits as the magnetic field is turned as in our measurements. The results are summarized in 
Figure~\ref{FA-d}.  The calculated angle dependence is also very weak up to 40$^{\circ}$. Due to
spin-orbit splitting, near 40$^{\circ}$-60$^{\circ}$, the calculated orbit frequency is suddenly
almost doubled with a resulted complicated orbit which involves both the cup surface and the interior of the
mug-shape surface. After this sudden increase, a new orbit is obtained as the interior of the mug-shape
surface (see Fig.~\ref{FA-d} which has, interestingly almost the same magnitude as original frequency. 
Hence, the overall angle dependence is very weak, as we found in our measurements. 
Because of the two Fermi surface near each other due to spin-orbit splitting and complicated
orbit shape, it is unlikely for the electrons to oscillate around such a complicated orbit
coherently and therefore one may
expect small intensity, which is consistent with decreasing intensity in our measurements.  

The angle dependence of SdH oscillation at ambient pressure maps out two 3D topological Fermi surfaces of two electron pockets from multiband model applied with LK fit in equation (\ref{LKfit}). From the magnetoresistance raw data at ambient pressure, we subtract the second order polynomial background. The FFT spectra shows three bands, $F_{\gamma}$ = 32.5 T, $F_{\alpha}$ = 240.5 T, $F_{\beta}$ = 258.0 T and their second and third order harmonic oscillations. The higher harmonic peaks of 
$F_{\alpha}$ and $F_{\beta}$ indicate high quality and homogeneity of the single crystal. The 1.8 K SdH oscillation data were fitted by the multiple bands of three dimensional LK formula and got the corresponding Berry phases, ($\phi_{\gamma}$ = $\pi$, $\phi_{\alpha}$ = 0.88$\pi$, and $\phi_{\beta}$ = 0.88$\pi$), indicating that T$_{d}$-MoTe$_{2}$ at ambient pressure is a possible three dimensional topological semimetal with 3D topological phase shift, $\delta = - \frac{1}{8}$ \cite{Lukyanchuk2004,Lukyanchuk2006,Champel2001} for electron pocket in equation (\ref{phase}).  
\clearpage

\section{Shubnikov-de Haas oscillations}
We provide the angle dependence of Shubnikov-de Haas oscillations for mapping the Fermi surface of MoTe$_2$ at ambient pressure in subsection~\ref{AQO}. The definitions of topological phase shift and nontrivial Berry's phase are derived from the Lifshitz-Onsager quantization relation. Applying global fitting of Lifshitz-Kosevich formula with bumps, we get uncertainty and correlation plots of the Dingle temperature and Berry's phase.  

\subsection{Berry's phase and topological phase shift}
The Lifshitz-Onsager quantization relation describes the closed trajectory of a charge carrier by an external magnetic field B as a function of Berry's phase,
\begin{align}
A_n\frac{\hbar}{eB}= 2\pi(n +\frac{1}{2}-\frac{\Phi_B}{2\pi}) = 2\pi (n+ \gamma).
\end{align}
Here, $A_n$ is the cross-sectional area of the Fermi surface related to the Landau level (LL) $n$ and Berry's phase $\Phi_B$. The periodic SdH oscillation in magnetoresistance follows the Lifshitz-Kosevich (LK) formula for three-dimensional system\cite{Shoenberg1984,Roth1966,Landwehr1991} as,
\begin{align}\label{LKfit}
A(B,T) &= \frac{\Delta\rho}{\rho_0} \propto (\frac{\hbar\omega_c}{E_F})^{1/2}\exp (-2\pi ^2 k_B T_D/\hbar \omega_c) \nonumber \\
& \frac{2\pi ^2 k_B T/\hbar \omega_c}{\sinh (2\pi ^2 k_B T/\hbar \omega_c)}.
\end{align}
Here, $\rho_0$ is the nonoscillatory component of resistivity at zero field, and other parameters are Dingle temperature $T_D$, cyclotron frequency $\omega_c =eB/m^*$, and Boltzmann's constant $k_B$. Applying the 3D LK formula, the effective mass of charge carrier m$^*$ could be extracted from the temperature dependence of oscillation amplitude. The longitudinal resistance in 3D system\cite{Shoenberg1984,Roth1966,Landwehr1991,Adams1959},
\begin{equation}\label{phase}
\rho_{xx} = \rho_0[1+ A(B,T)\cos 2\pi(B_F/B-\delta+\gamma)],
\end{equation}
$1/B_F$ is the SdH frequency, and $\delta $ is the topological phase shift, which is determined by the dimensionality of Fermi surface, $\delta = 0$ (or $\delta = \pm \frac{1}{8}$) for the 2D (or 3D) system\cite{Lukyanchuk2004,Lukyanchuk2006,Champel2001}. Therefore, $|\gamma - \delta|= |1/2-\phi_B/2\pi-\delta|$ between 0 and 1/8 indicates a nontrivial $\pi$ Berry's phase. The detail of band structure such as frequency, Dingle temperature, effective mass and Berry's phase are listed in table~\ref{fitting}. Berry's phase determination is very sensitive to the Zeeman effect and the Fermi surface might distort close to the quantum limit. The fitting of Berry's phase in this paper focuses on relatively lower fields below 18 T. The raw SdH oscillation data curve is fitted with 3D multiband LK formula and uncertainty analysis from a Bayesian perspective with $Bumps$\cite{bumps, bumps2} including Markov Monte Carlo method \cite{bumps, bumps2} for the joint distribution of parameter probability.  Table~\ref{fittingpara} indicates oscillation frequency (proportional to cross section area of Fermi surface), Dingle temperature, effective mass of each pocket, the corresponding Berry's phase from ambient pressure to 1.8 GPa.
\\

\subsection{Global fitting- bumps}
To find the global minimum of multiple bands model in Lifshitz-Kosevich formula, we get the expectation values of effective mass and oscillation frequency of the charge carriers according to the temperature dependence of oscillation amplitude and fast Fourier Transform of Shubnikov-de Haas oscillations. First, we apply gradient descent to quickly search local minimum and include the real experimental errors for global fitting of LK formula. The consistent results of variables such as Dingle temperature and Berry's phase indicate convergence of the fitting parameters. Bumps is a set of free and public routines for complicated curve fitting, uncertainty analysis and correlation analysis from a Bayesian perspective\cite{bumps, bumps2}. To see the distributions of uncertainty and correlation plots, We run bumps for the the rest of variables, amplitude of oscillation, Dingle temperature and Berry phase. In general the faster algorithms (Levenberg-Marquardt, Quasi-Newton BFGS) tend to find the local minimum quickly rather than the slower global minimum. Bumps provides uncertainty analysis which explores all viable minima and finds confidence intervals on the parameters based on uncertainty from experimental errors. 
\\

\begin{figure}
\centering
\subfigure{\includegraphics[scale=0.26, trim= 170 70 120 120,clip]{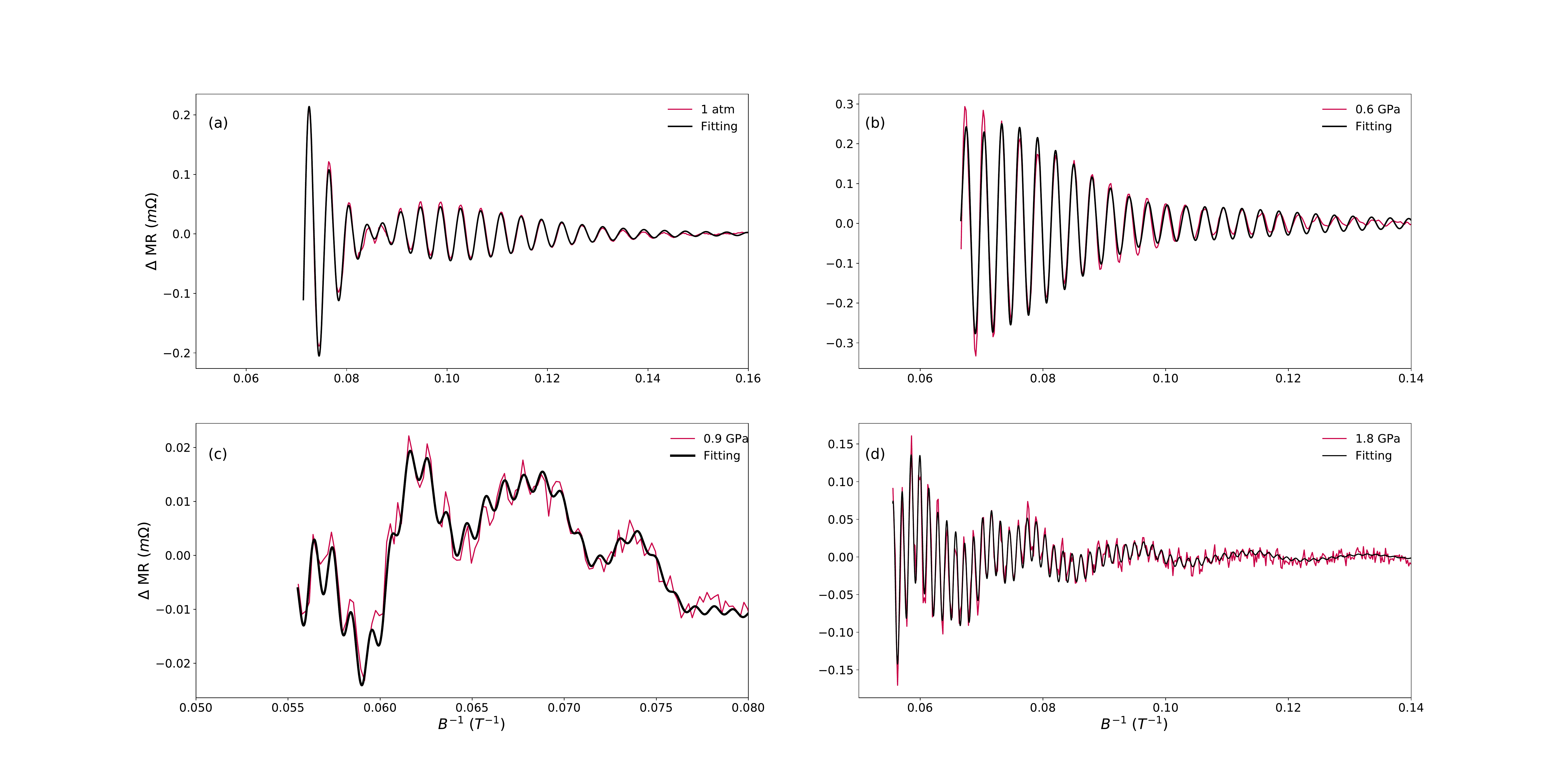}}
  \caption{\label{Berry2}LK fits in multiple band system in MoTe$_2$ made at 1.8 K and (a) 1atm through Physical Property Measurement System (PPMS), and (b) 0.6 GPa one is made in a dilution refrigerator were taken using a lakeshore LS370 AC resistance bridge down to 0.1 K. Higher pressure cases such as (c) 0.9 GPa at 0.27 K and (d) 1.8 GPa at 0.3 K are preformed by Oxford Heliox. 
  }
\end{figure}

\begin{figure}
\centering
\subfigure{\includegraphics[scale=0.7,trim= 95 120 45 142,clip]{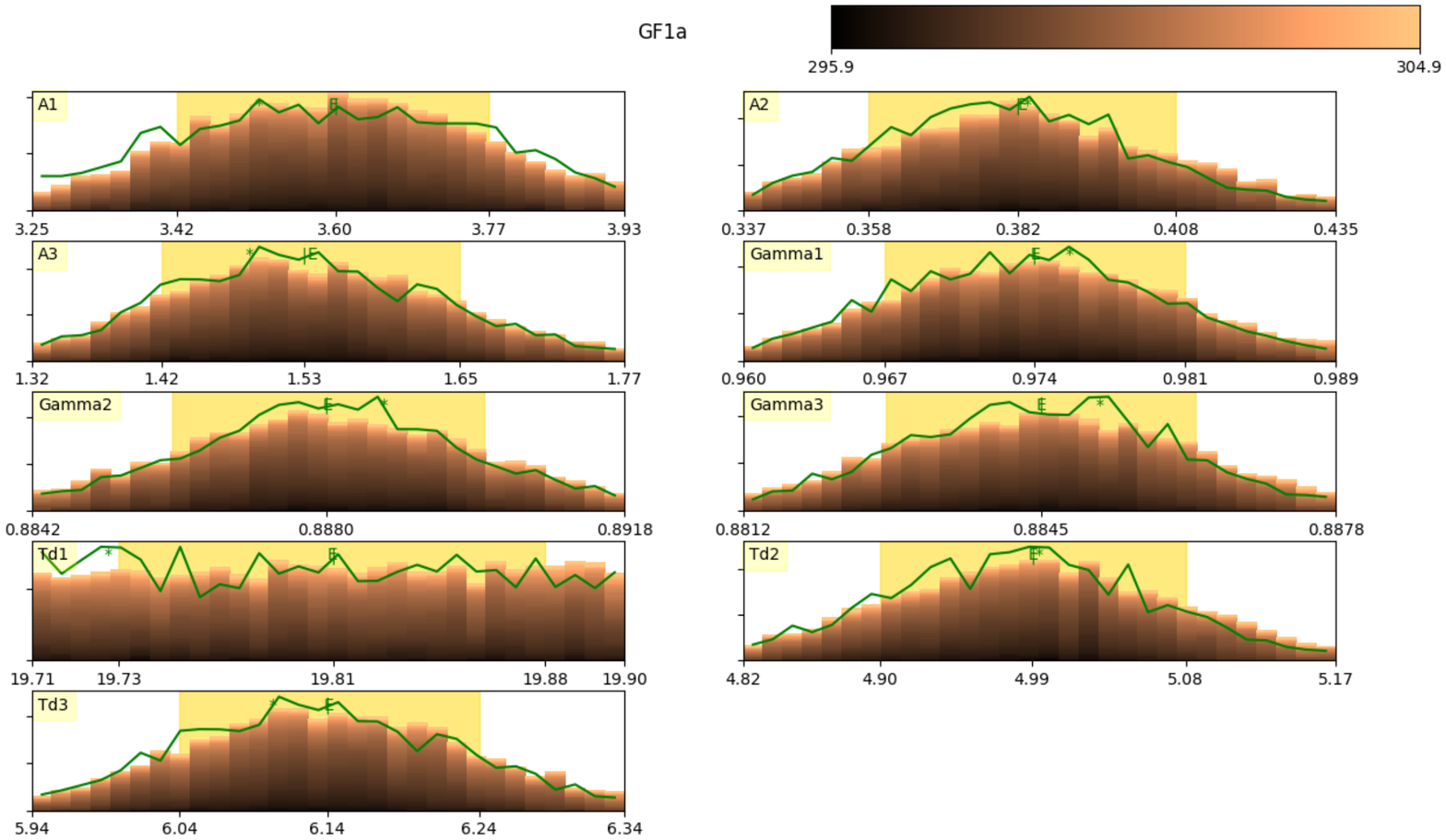}}
\subfigure{\label{1aCor}\includegraphics[scale=0.95, trim= 90 240 80 255,clip]{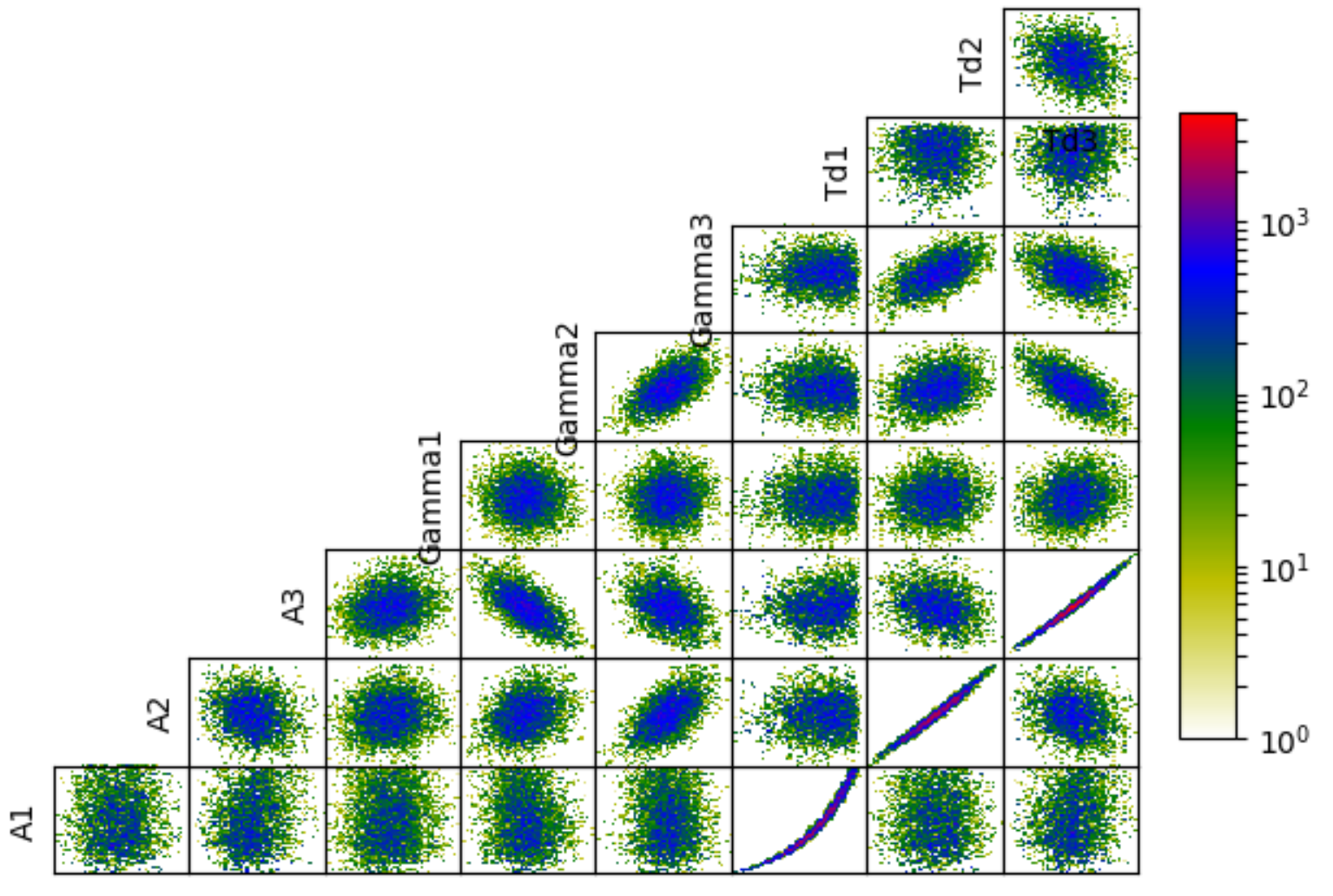}}
  \caption{ \label{1aU}This is the (a) uncertainty of Lifshitz-Kosevich formula fitting in MoTe$_2$ made at 1.8 K and 1 atm. The Berry's phase of F$_{\gamma}$, F$_{\alpha}$, and F$_{\beta}$ are (0.97 $\pm$ 0.02)$\pi$, (0.89 $\pm$ 0.01)$\pi$ and (0.88 $\pm$ 0.01)$\pi$. The parameters, B, ${\gamma}$, T$_d$, and $A$  are oscillation frequency, Berry's phase, Dingle temperature and maximum oscillation amplitude. The number aligns from the smaller to larger cross section area, oscillation frequency. (b) The 2D correlation plots between each two parameters in our fitting formulas. There is positive correlation between maximum oscillation amplitude and Dingle temperature, which they all tune the oscillation damping factor, the damping rate. The Berry's phase shows no correlation with Dingle temperature and oscillation amplitude.}
\end{figure}

Bumps includes Markov chain Monte Carlo (MCMC) methods \cite{bumps, bumps2} to compute the joint distribution of parameter probabilities. The MCMC explores the space using a random walk and requires hundreds of thousands of function evolutions to explore the search space. The histogram range represents the 95 $\%$ credible interval, and the shaded region represents the 68 $\%$ credible interval. For full uncertainty analysis, bumps uses a random walk to explore the viable parameter space near the minimum, showing pair-wise correlations between the different parameter values. The 2D correlation plots indicate the correlation relationship between multiple parameters in the fitting function. With Bumps, we could check the convergence of fitting sequence and compare different local minimum to get the global minima. Figure~\ref{Berry2} shows the best fit curves of SdH oscillation signal from 1 $atm$ to 1.8 GPa. We show the expectation value and uncertainties for fitting data under 1atm (Fig.~\ref{1aU}), 0.6 GPa (Fig.~\ref{p6rU}), 0.9 GPa (Fig.~\ref{1rU}) and 1.8 GPa (Fig.~\ref{18rU}).
\\

\begin{table}[]
\label{fitting}
\begin{tabular}{|c|c|c|c|c|}
\hline
P (GPa)                  & F (T)  & $T_D$ (K) & $m^*$ ($m_e$) & $\phi$ ($\pi$) \\ \hline
\multicolumn{5}{|c|}{}                                          \\ \hline
\multirow{3}{*}{$\sim$0} & 32.50  & 19.81  & 0.31    & 0.97     \\ \cline{2-5} 
                         & 240.50 & 4.99   & 0.53    & 0.89     \\ \cline{2-5} 
                         & 258.00 & 6.14   & 0.59    & 0.88     \\ \hline
\multicolumn{5}{|c|}{}                                          \\ \hline
\multirow{3}{*}{0.6}     & 35.00  & 18.20  & 0.30    & 0.95     \\ \cline{2-5} 
                         & 330.30 & 4.16   & 0.71    & 0.89     \\ \cline{2-5} 
                         & 355.00 & 9.16   & 0.63    & 0.88     \\ \hline
\multicolumn{5}{|c|}{}                                          \\ \hline
\multirow{4}{*}{0.9}     & 32.00  & 7,91   & 0.21    & 0.87     \\ \cline{2-5} 
                         & 170.00 & 5.12   & 1.02    & 0.93     \\ \cline{2-5} 
                         & 245.00 & 2.97   & 1.08    & 0.88     \\ \cline{2-5} 
                         & 966.00 & 3.54   & 1.75    & 1.02     \\ \hline
\multicolumn{5}{|c|}{}                                          \\ \hline
\multirow{4}{*}{1.8}     & 53.00  & 3.48   & 0.47    & 0.98     \\ \cline{2-5} 
                         & 110.00 & 8.48   & 0.57    & 0.77     \\ \cline{2-5} 
                         & 158.00 & 6.88   & 0.81    & 1.20     \\ \cline{2-5} 
                         & 690.00 & 3.94   & 0.97    & 1.10     \\ \hline
\end{tabular}
\caption{\label{fittingpara}The fitting parameters of LK formula in MoTe$_2$ under pressure P. We obtained oscillation frequency F from FFT and effective mass m$^*$ by the temperature dependence of FFT. 
Here, the sum of Berry's phase and phase shift $\phi$ ( phase shift determined by the dimensionality of Fermi surface, $\delta = 0$ or $\delta = \pm \frac{1}{8}$ for 2D or 3D ) could be directly gained and derived from LK formula.}
\end{table}

\begin{figure}
\centering
\subfigure{\label{p6rU}\includegraphics[scale=0.7, trim= 95 140 45 142,clip]{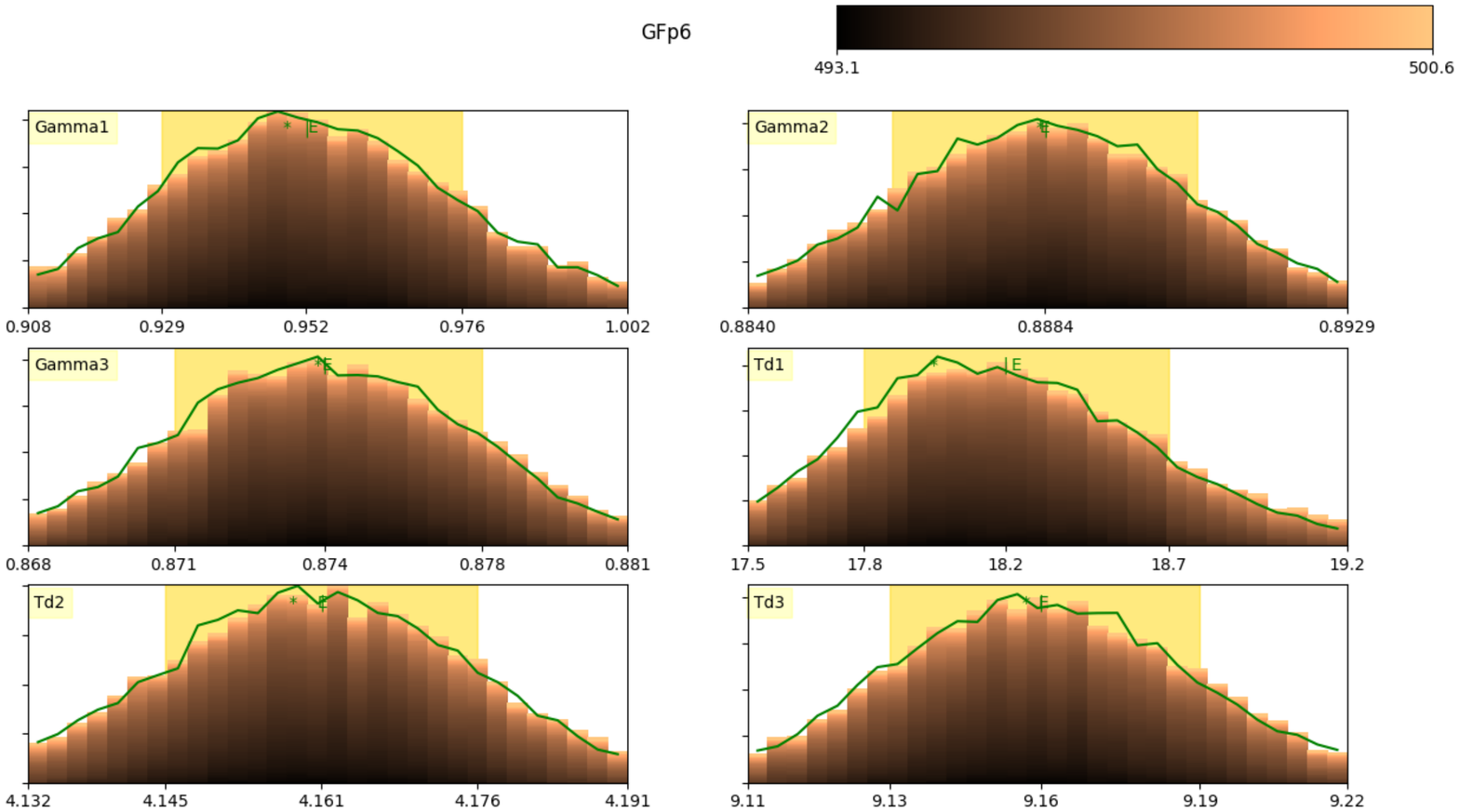}}
\subfigure{\label{p6rCor}\includegraphics[scale=0.9, trim= 90 240 80 255,clip]{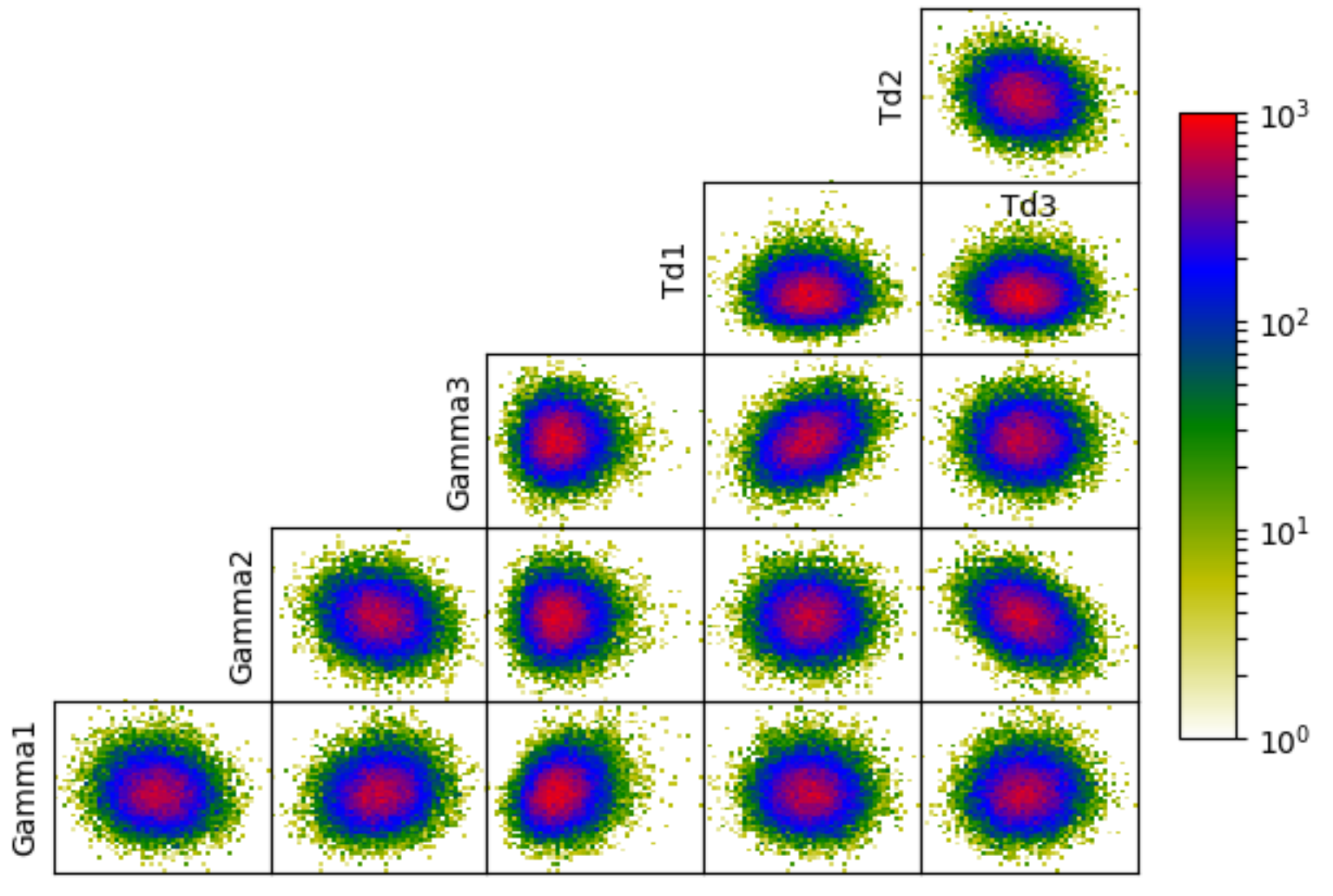}}
  \caption{\label{p6rU}This is the (a) uncertainty of Lifshitz-Kosevich formula fitting in MoTe$_2$ made at 0.1 K and 0.6 GPa. The Berry's phase of F$_{\gamma}$, F$_{\alpha}$, and F$_{\beta}$ are (0.95 $\pm$ 0.05)$\pi$, (0.89 $\pm$ 0.01)$\pi$ and (0.87 $\pm$ 0.01)$\pi$. The parameters, B, ${\gamma}$, and T$_d$ are oscillation frequency, Berry's phase, and Dingle temperature. The number aligns from the smaller to larger cross section area, oscillation frequency. (b) The 2D correlation plots between each two parameters in our fitting formulas. The Berry's phase shows no correlation with Dingle temperature. }
\end{figure}

\begin{figure}
\centering
\subfigure{\label{1rU}\includegraphics[scale=0.7, trim= 95 140 45 142,clip]{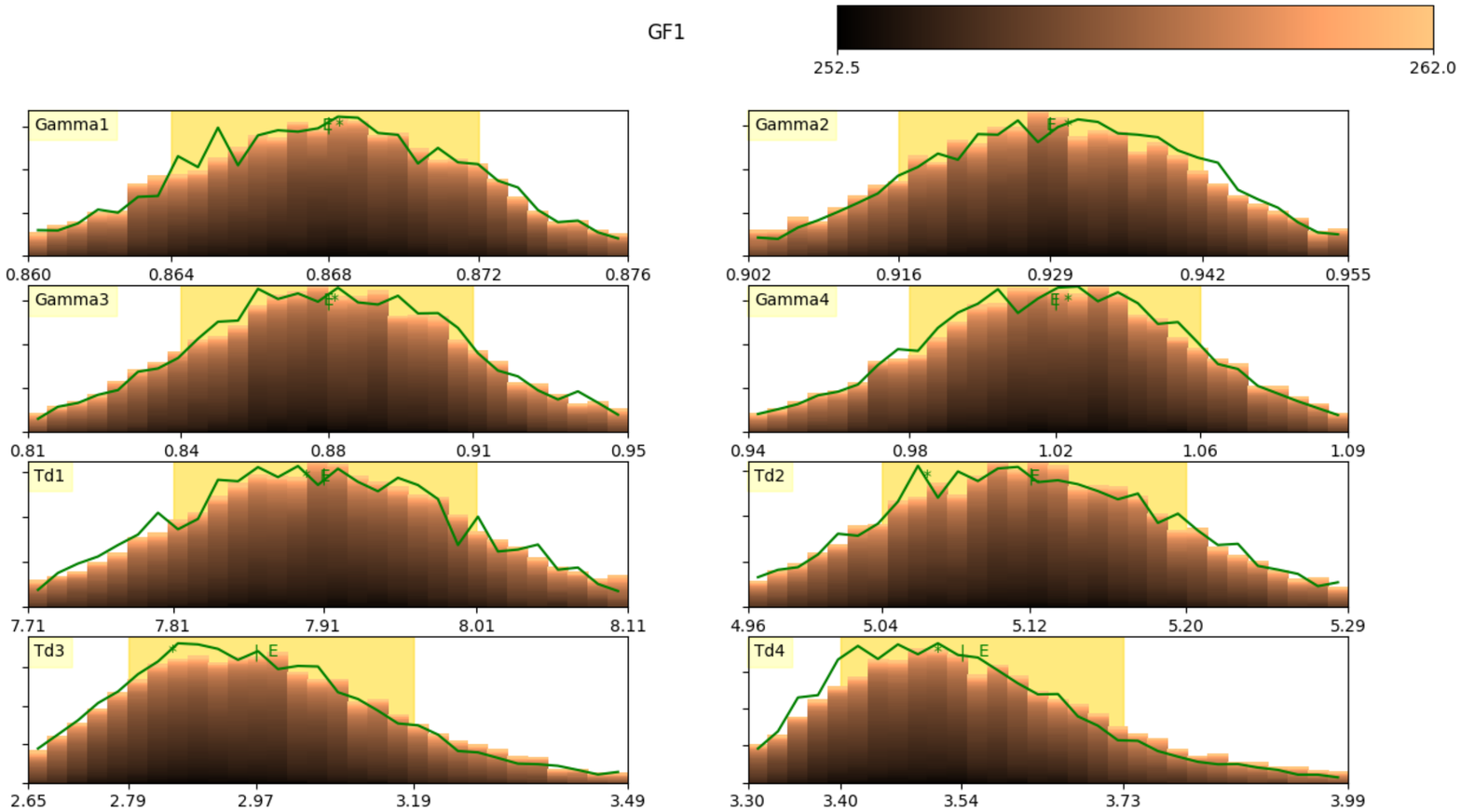}}
\subfigure{\label{1rCor}\includegraphics[scale=0.9, trim= 90 240 80 255,clip]{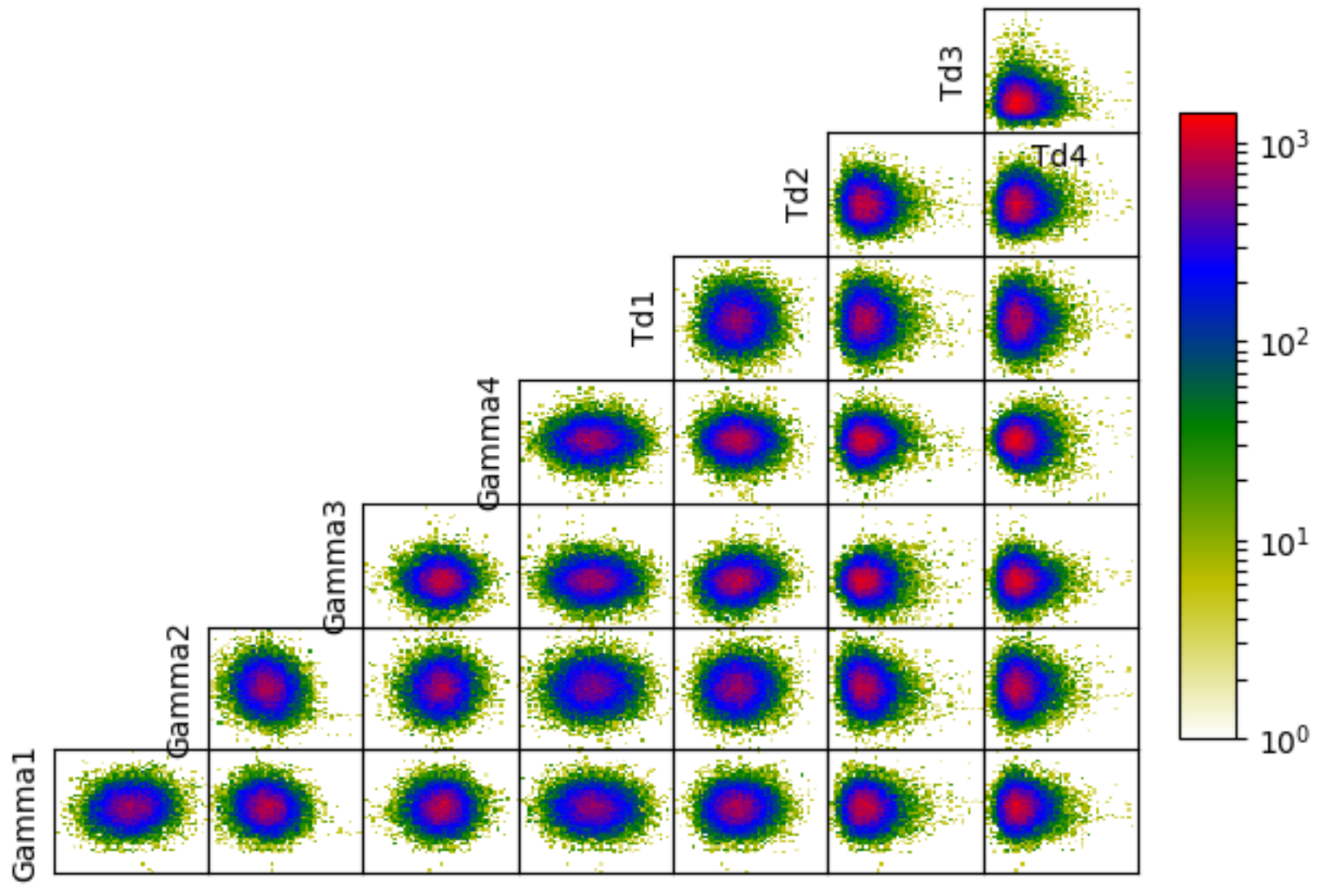}}
  \caption{\label{1rU}This is the (a) uncertainty of Lifshitz-Kosevich formula fitting in MoTe$_2$ made at 0.27 K and 0.9 GPa. The Berry's phase of F$_{\lambda}$, F$_{\mu}$, F$_{\nu}$, and F$_{\delta}$ are (0.87 $\pm$ 0.01)$\pi$, (0.93 $\pm$ 0.02)$\pi$, (0.88 $\pm$ 0.07)$\pi$, and (1.02 $\pm$ 0.08)$\pi$. The parameters, B, ${\gamma}$, and T$_d$ are oscillation frequency, Berry's phase, and Dingle temperature. The number aligns from the smaller to larger cross section area, oscillation frequency. (b) The 2D correlation plots between each two parameters in our fitting formulas. The Berry's phase shows no correlation with Dingle temperature.}
\end{figure}


\begin{figure}
\centering
\subfigure{\label{18rU}\includegraphics[scale=0.7, trim= 95 140 45 142,clip]{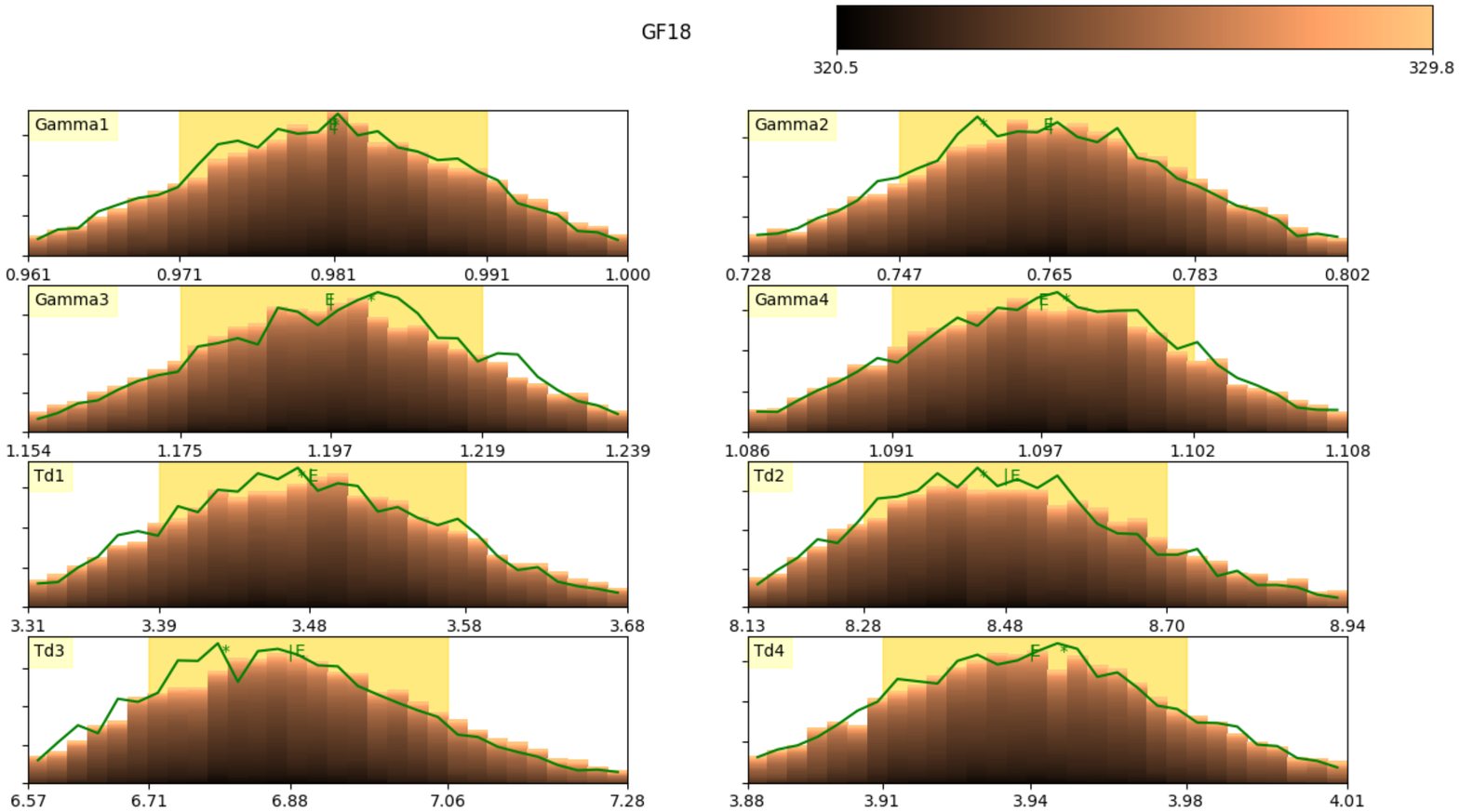}}
\subfigure{\label{18rCor}\includegraphics[scale=0.9, trim= 90 240 80 255,clip]{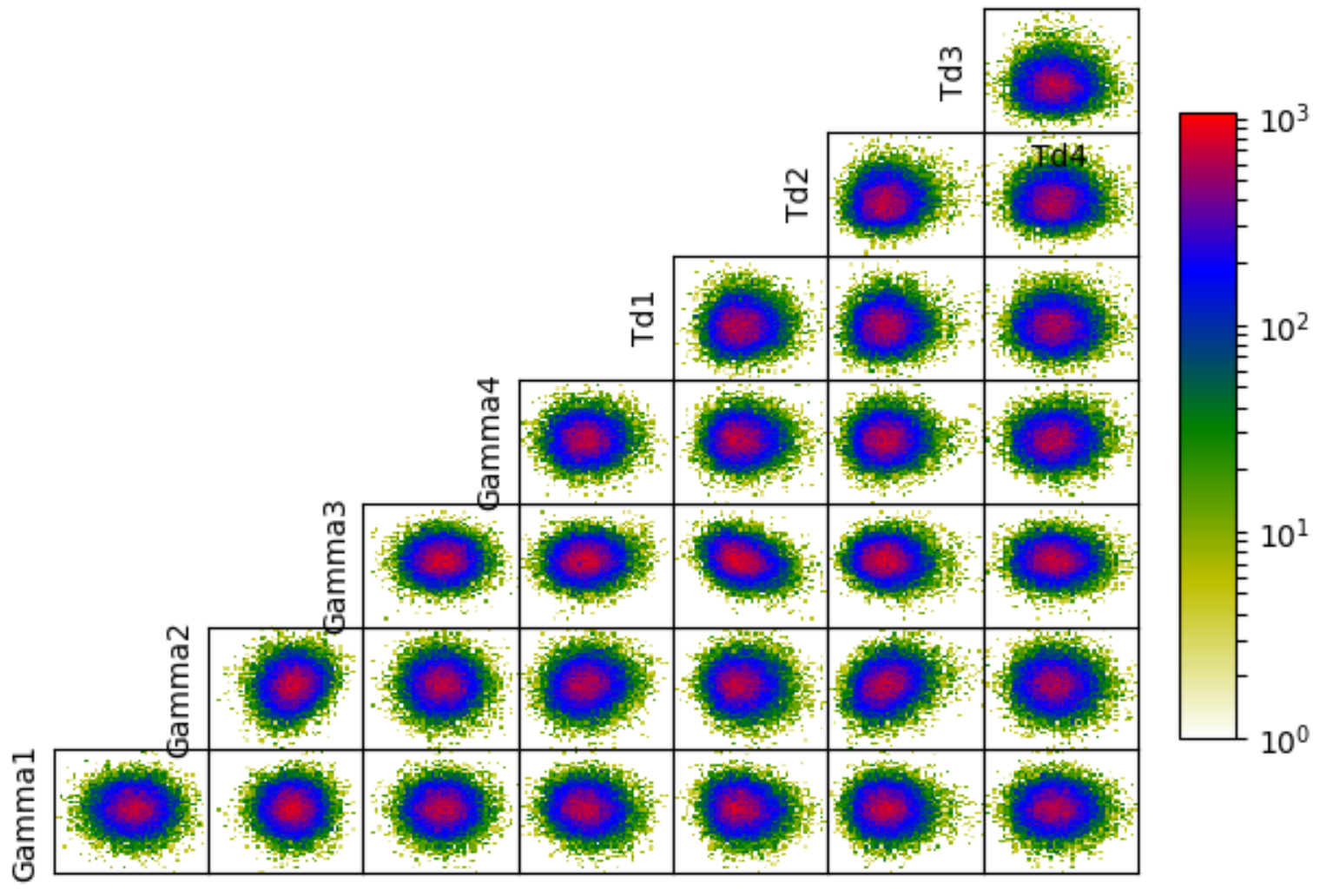}}
  \caption{\label{18rU}This is the (a) uncertainty of Lifshitz-Kosevich formula fitting in MoTe$_2$ made at 0.3 K and 1.8 GPa. The Berry's phase of F$_{\lambda}$, F$_{\mu}$, F$_{\nu}$, and F$_{\delta}$ are (0.98 $\pm$ 0.02)$\pi$, (0.77 $\pm$ 0.03)$\pi$, (1.20 $\pm$ 0.03)$\pi$, and (1.10 $\pm$ 0.01)$\pi$. The parameters, B, ${\gamma}$, and T$_d$ are oscillation frequency, Berry's phase, and Dingle temperature. The number aligns from the smaller to larger cross section area, oscillation frequency. (b) The 2D correlation plots between each two parameters in our fitting formulas. The Berry's phase shows no correlation with Dingle temperature.}
\end{figure}

\begin{figure}
\centering
\subfigure{\label{p8}\includegraphics[scale=0.3, trim= 40 40 60 30,clip]{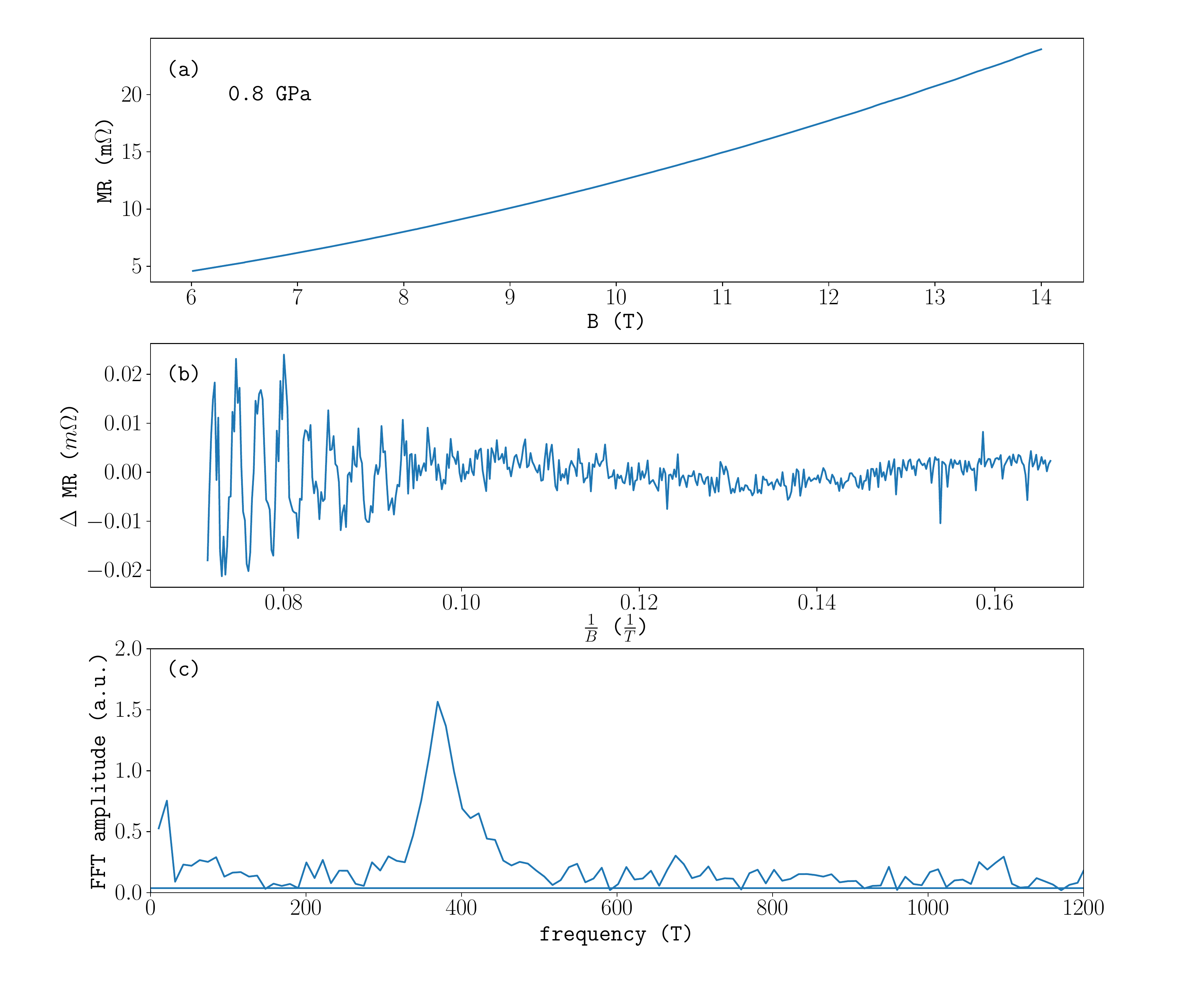}}
  \caption{(a) The longitudinal MR, (b) SdH oscillations and (c) its FFT of the bulk MoTe$_2$ sample measured at 1.8 K and 0.8 GPa with magnetic field parallel to c axis.}
\end{figure}
\begin{figure}
\centering
\subfigure{\label{1p1}\includegraphics[scale=0.3, trim= 40 40 60 20,clip]{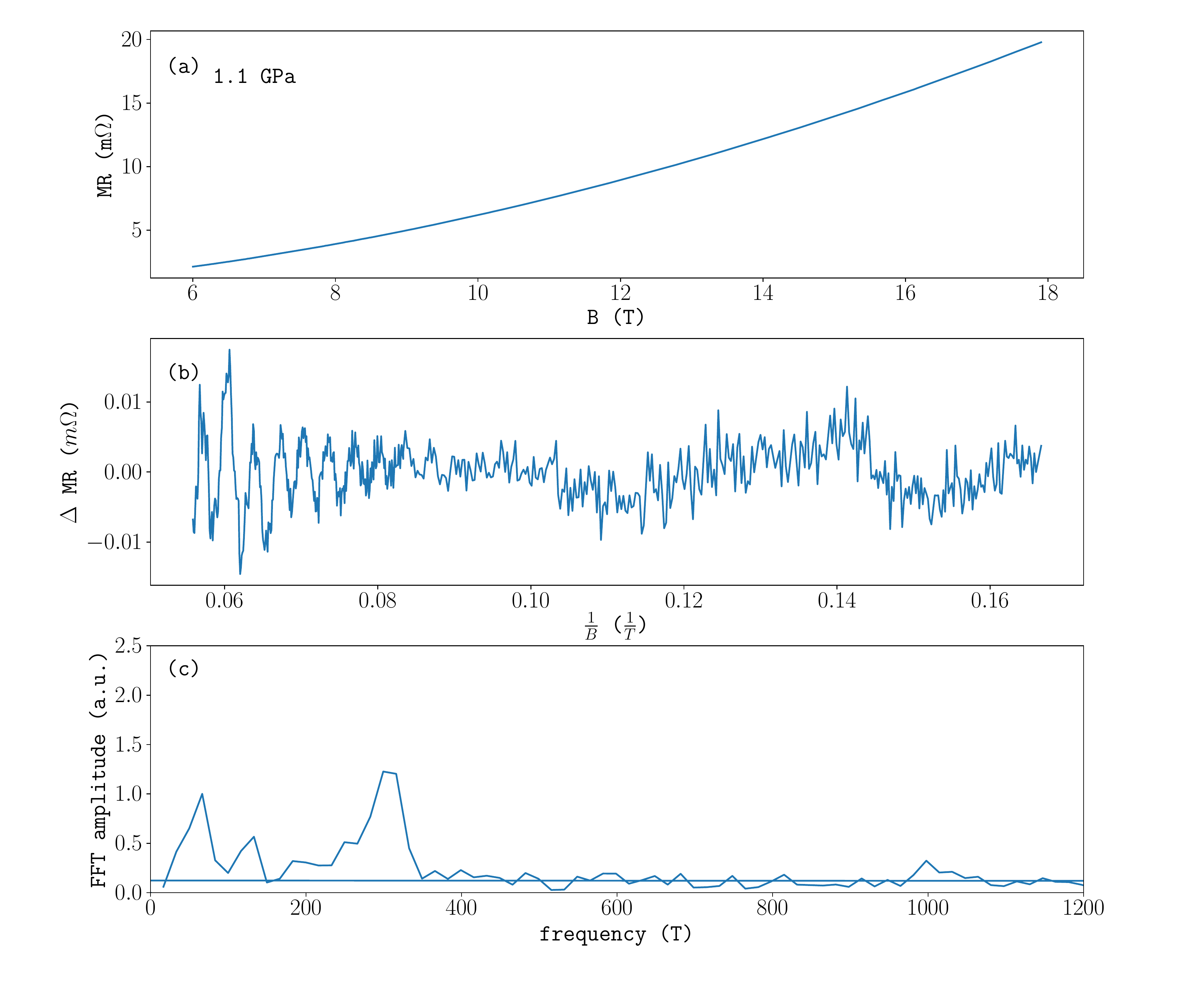}}
  \caption{(a) The longitudinal MR, (b) SdH oscillations and (c) its FFT of the bulk MoTe$_2$ sample measured at 0.3 K and 1.1 GPa with magnetic field parallel to c axis.}
\end{figure}
\begin{figure}
\centering
\subfigure{\label{1p2}\includegraphics[scale=0.3, trim= 40 40 60 25,clip]{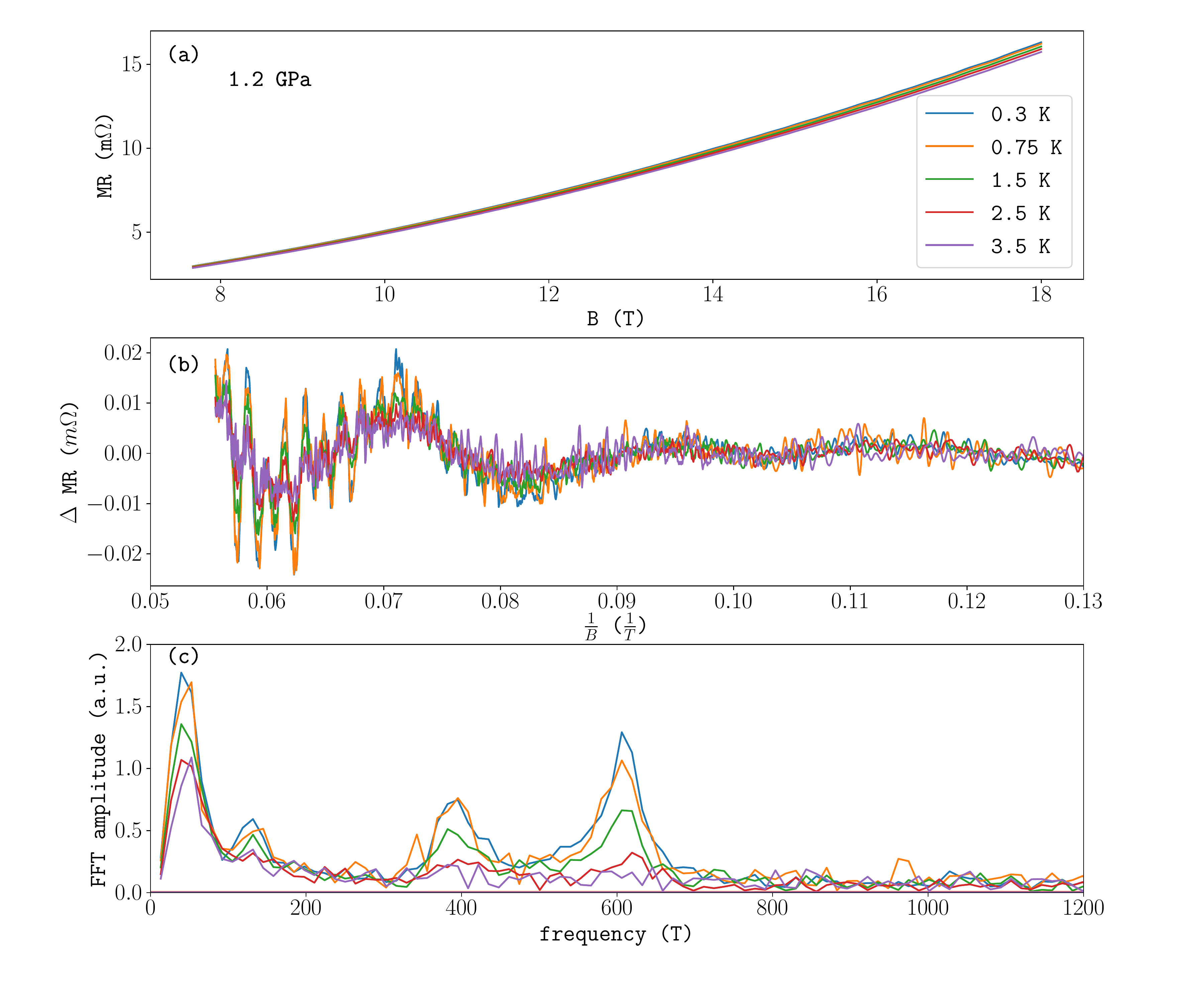}}
  \caption{(a) The longitudinal MR, (b) SdH oscillations and (c) its FFT of the bulk MoTe$_2$ sample measured at 1.8 K and 1.2 GPa with magnetic field parallel to c axis.}
\end{figure}

